\documentclass[11pt]{article}

\usepackage[skip=2pt,labelfont=normalsize,textfont=footnotesize]{caption}
\usepackage[top=1in, bottom=1in, left=1in, right=1in]{geometry}
\usepackage{setspace}
\usepackage{array} 
\usepackage[normalem]{ulem}
\usepackage{graphicx,psfrag}
\graphicspath{{diagrams/}}
\usepackage{subfigure}
\usepackage{pst-pdf}  

\usepackage{float}
\usepackage{amsmath}
\usepackage{amssymb}
\usepackage{bm}
\usepackage{chemarr}
\usepackage{url}
\usepackage{balance}
\usepackage{enumerate}
\usepackage{color}
\usepackage[utf8]{inputenc}
\usepackage{verbatim}
\usepackage{float}
\usepackage{xr}

\usepackage[fancythm,fancybb,draft,elsart]{jphmacros2e} 

\usepackage[norefs,nocites]{refcheck} 
\usepackage[numbers,sort&compress,sectionbib]{natbib} 
\usepackage[colorlinks=true,linkcolor=blue]{hyperref}

\newcommand{\zrxn}[2]{\varnothing \xrightarrow{#1} #2}
\newcommand{\orxn}[3]{#1 \xrightarrow{#2} #3}
\newcommand{\trxn}[4]{#1 + #2   \xrightarrow{#3}  #4}

\newcommand{\orxna}[3]{#1 &\xrightarrow{#2} #3}

\newcommand{\cX}[1]{\left[#1\right]}
\newcommand{\dotcX}[1]{\dot{\left[#1\right]}}
\newcommand{\ms}{\mathcal{S}}
\newcommand{\mr}{\mathcal{R}}
\newcommand{\mv}{\mathcal{M}}
\newcommand{\tu}[1]{\textup{#1}}
\newcommand{\koff}{k^{\textup{off}}}
\newcommand{\kon}{k^{\textup{on}}}

\newcommand{\ptot}{P^{\textup{tot}}}

\newcommand{\bndxy}[2]{#1\textup{:}#2}
\newcommand{\gene}[1]{\mathcal{G}_{#1}}

\newcommand{\bbeta}{\bar{\beta}}
\newcommand{\kf}{k^{\tu{f}}}
\newcommand{\kr}{k^{\tu{r}}}
\newcommand{\kcat}{k^{\tu{cat}}}
\newcommand{\etot}{E^{\tu{tot}}}
\newcommand{\kbarr}{\bar{k}^\mrm{r}}
\newcommand{\kbarf}{\bar{k}^\mrm{f}}
\newcommand{\albar}{\bar{\alpha}}
\newcommand{\Kd}{K^\mrm{d}}
\newcommand{\draftbreak}{}

\title{Modular decomposition and analysis of biological networks}

\author{Hari~Sivakumar\thanks{
           Corresponding author.  Address: 
           Electrical and Computer Engineering,
	   University of California,
	   Santa Barbara, CA~93106-9560, U.S.A.,
	   Tel.:~(805)893-7785} \\
	Electrical and Computer Engineering, \\
	University of California Santa Barbara, Santa Barbara, CA,\\
	\and Stephen R. Proulx\\
	Ecology, Evolution and Molecular Biology,\\
	University of California Santa Barbara, Santa Barbara, CA,
	\and Jo{\~a}o P. Hespanha \\
	Electrical and Computer Engineering, \\
	University of California Santa Barbara, Santa Barbara, CA}
\date{}

\newtheorem{rules}{Rule}

\begin{document}
\maketitle

\begin{abstract}
 This paper addresses the decomposition of biochemical networks into
  functional modules that preserve their dynamic properties upon
  interconnection with other modules, which permits the inference of
  network behavior from the properties of its constituent modules. The
  modular decomposition method developed here also has the property
  that any changes in the parameters of a chemical reaction only
  affect the dynamics of a single module. To illustrate our results,
  we define and analyze a few key biological modules that arise in
  gene regulation, enzymatic networks, and signaling pathways. We also
  provide a collection of examples that demonstrate how the behavior
  of a biological network can be deduced from the properties of its
  constituent modules, based on results from control systems theory.
\end{abstract}

\bigskip


\clearpage

\section{Introduction}
\label{sec:intro}
Network science and control systems theory have played a significant role in the advancement of systems and synthetic biology in recent years. Tools
from graph theory and system identification have been used to
understand the basic subunits or ``motifs'' in biological networks
\cite{milo2002network,alon2007network}, and to identify the existence
(or non-existence) of pathways in large networks
\cite{bruggeman2002modular,Kholodenko1,Kholodenko2}. With a deeper
understanding of mechanisms that exist in biology, synthetic biologists have used bottom-up construction techniques to engineer
new biological networks, and modify or optimize the behavior of
existing networks \cite{endy2005foundations,biobrick}, with impressive results
\cite{rothemund2006folding,Elowitz2000aa,gardner2000construction}.

\medskip

\subsection*{Modularity}
One of the key ideas that has revolutionized synthetic and systems
biology is the conception of a biological network as a collection of
functionally isolated interacting components. To the authors'
knowledge, Hartwell \textit{et al.}~were among the first to suggest
that this idea could reduce the complexity of analyzing these networks
\cite{Hartwell99}. For example, properties like stability and
robustness can be predicted just from properties of each individual
component in the network and knowledge of the interconnection
structure. From a computational
perspective, computing network parameters such as its equilibrium
point(s) can be greatly simplified, since the computations can be done
over a set of components as opposed to over the entire
network. From an evolutionary standpoint,
grouping a network into components is useful to analyze the
evolvability of each component. Finally, for experimental
identification purposes, delimiting a network into components reduces
the complexity of the estimation problem and often requires less
experimental data.

\medskip

It is generally understood that to be useful, a biological component
needs to exhibit a property broadly defined as \emph{modularity}, but
there is little consensus on its definition and on how it is perceived
\cite{Sauro08}. Synthetic biologists and mathematical biologists
postulate that a biological component exhibits
modularity if its dynamic characteristics remain the same before and
after interconnection with other components \cite{Sauro08}. However,
synthetic biologists sometimes argue that modularity is \emph{not} an
inherent property of biological network components, because often
components have dynamics that are susceptible to change upon
interconnection with other components. This
has been termed many things, including ``hidden feedback''
\cite{ventura2008hidden}, ``retroactivity'' \cite{DelVecchio08} and
``loading effects'' \cite{sontag2011modularity}, and is akin to an
electrical component whose output voltage changes upon the addition of
a load. To obviate this lack of modularity and to effectively isolate
two or more synthetic components from each other, insulation
components have been designed \cite{DelVecchio08,del2008retroactivity}
and have proven to be useful \cite{franco2011timing}.

\medskip

Mathematical biologists on the other hand, believe that the fact that
some synthetic components can be subject to loading effects upon
interconnection with other components provides no reason to believe
that biological networks cannot be analytically delimited into
components that exhibit modularity.  In this field, a biological
network is generally represented by a system of ordinary differential
equations (ODEs). Every one of these equations is then assigned to a
component, with appropriate input-output relationships to ensure that
the composition of all components will reconstruct the original system
of ODEs. This method of decomposing a biological network has
proven useful in deriving results pertaining to the existence of
multiple equilibrium points in a network of interconnected components
\cite{angeli2003monotone,angeli2004interconnections,angeli2004multi},
and the stability of these points
\cite{arcak2006diagonal,arcak2008passivity}. A
potentially undesirable feature of the approaches followed in the
above-mentioned references is that, while the components considered
were dynamically isolated from each other, the parameters of a
particular chemical reaction in the network (such as the
stoichiometric coefficients or the rate constants) could appear in
more than one component. Consequently, a change in a single chemical
reaction could result in several distinct components changing their
internal dynamics.

\medskip

Evolutionary biologists themselves have different perspectives on modularity. One is that modularity
is an inherent property of biological network components that enables
them to evolve independently from the rest of the network in response
to shock or stress, hence enhancing future
evolvability. \cite{espinosa2010specialization,clune2013evolutionary}. Another perspective is that natural selection occurs modularly, and that
this selection preserves certain properties of a biological network by
allowing individual aspects of a component to be adjusted without
negative effects on other aspects of the phenotype. Whichever
definition is used, evolutionary biologists rely heavily on biological
network parameters being associated with a unique component. For
example, parameters that are "internal" to a functional biological
component, but that do not affect significantly the input-output
behavior of the component, are considered as neutral traits
\cite{proulx2010standard}, meaning that their values can change
because of genetic drift.

\subsection*{Two notions of modularity}
In this paper, we attempt to unify the notion of modularity in the
context of biological networks, using an analytical approach. We say
that a biological component is a \emph{module} if it admits both
\emph{dynamic modularity} and \emph{parametric modularity}. The former
implies that the properties of each module do not change upon
interconnection with other modules, and the latter implies that the
network parameters within a module appear in no other module. In this
sense, a synthetic component that undergoes loading effects upon
interconnection with other components does \emph{not} exhibit dynamic
modularity, and a component whose internal dynamics depend on
parameters that also affect other components does \emph{not} exhibit
parametric modularity.

\medskip

Dynamic modularity is essential to infer the behavior of a biological
network from the behavior of its constituent parts, which can enable
the analyses of large networks and also the design of novel
networks. Parametric modularity, which has not been explicitly
mentioned as often in the literature, is a useful property when
identifying parameters in a biological network, like in
\cite{bruggeman2002modular,Kholodenko1,Kholodenko2}, and also for
evolutionary analysis of network modules.

\medskip
\subsection*{Novel contributions}
A key contribution of this paper is the development of a systematic
method to decompose an arbitrary network of biochemical reactions into
modules that exhibit both dynamic and parametric modularity. This
method is explained in detail in Section~\ref{sec:probstat} and is based
on three rules that specify how to partition species and reactions
into modules and also how to define the signals that connect the
modules. An important novelty of our approach towards a modular
decomposition is the use of reaction rates as the communicating signals
between modules (as opposed to species concentrations). We show that aside from permitting parametric
modularity, this allows the use of summation junctions to combine
alternative pathways that are used to produce or degrade particular
species.

\medskip

To illustrate the use of our approach, in Section~\ref{sec:examples}
we introduce some key biological modules that arise in gene regulatory
networks, enzymatic networks, and signaling pathways. Several of these
biological systems have been previously regarded in the literature as
biological "modules", but the "modules" proposed before did not
simultaneously satisfy the dynamic and parametric modularity
properties. We then analyze these modules from a systems theory
perspective. Specifically, we introduce the Input-output static
characteristic function (IOSCF) and the Linearized Transfer Function
(LTF), and explain how these functions can help us characterize a
module by properties such as monotonicity and stability.  \medskip

Section~\ref{sec:negfdbk} is devoted to demonstrating how one can predict the behavior of large networks from
basic properties of its constituent modules. To this effect, we review
three key basic mechanisms that can be used to combine simple modules
to obtain arbitrarily complex networks: cascade, parallel, and
feedback interconnections. We then illustrate how modular
decomposition can be used to show that the Covalent Modification
network \cite{goldbeter1981amplified} can only admit a single stable
equilibrium point no matter how the parameters of
the network are chosen, and to find parameter regions where the
Repressilator network \cite{Elowitz2000aa} converges to a stable
steady-state.

\medskip

\section{Modularity and Decomposition}
\label{sec:probstat}

From this paper's perspective, a biological network is viewed as a
collection of elementary chemical reactions, whose dynamics are
obtained using the law of mass action kinetics.  Our goal is to
decompose such a network into a collection of interacting
\emph{modules}, each a dynamical system with inputs and outputs, so
that the dynamics of the overall network can be obtained by
appropriate connections of the inputs and outputs of the different
modules that comprise the network. It is our expectation that the
decomposition of a biological network into modules will reduce the
complexity of analyzing network dynamics, provide understanding on the
role that each chemical species plays in the function of the network,
and permit accurate predictions regarding the change of behavior that
would arise from specific changes to components of the network.

\medskip

For a network to be truly modular, we argue that the different modules
obtained from such a decomposition should exhibit both \emph{dynamic
modularity} and \emph{parametric modularity}. A module exhibits dynamic modularity if its dynamics remain unchanged upon interconnection with other modules in the network, and admits parametric modularity if the parameters associated with that module appear in no other module of the network. Consequently, changing the parameters of a
single chemical reaction affects the dynamics of only a single module. 

\medskip

In the rest of this section, we develop rules that express conditions
for a module to admit dynamic and parametric
modularity. To express these rules, we introduce a
graphical representation that facilitates working with large
biological networks.

\subsection{Representing a biological network}
\label{subsec:rep}

A biological network can be represented in multiple ways, including as a
system of ordinary differential equations (ODEs) associated with the
law of mass action kinetics, a directed bipartite species-reaction
graph (DBSR), or a dynamic DBSR graph that we introduce below.

\subsubsection*{Mass Action Kinetics (MAK) Ordinary Differential Equations (ODEs)}

A set of species involved in chemical reactions can be expressed as a
system of ODEs using the law of mass
action kinetics (MAK) when the species are well-mixed and their copy
numbers are sufficiently large. For a network involving the species
$\ms_j$, $\forall j\in\{1,2,\dots,N_\mrm{species}\}$ and the reactions
$\mr_i$, $\forall i\in\{1,2,\dots,N_{\mrm{reactions}}\}$, the MAK
results in a system of ODEs whose states are the concentrations
$\cX{\ms_j}$, $\forall j$ of the different chemical species; the
ODE representing the dynamics of a specific species $\ms_j$ is given
by
\begin{align}\label{eq:MAK}
  \dot{\cX{\ms_j}} = \sum\limits_{i=1}^{N_{\mrm{reactions}}} \psi_{ji}(k_1,k_2,\dots,\cX{\ms_1},\cX{\ms_2},\dots)
\end{align}
where $\psi_{ji}(k_1,k_2,\dots,\cX{\ms_1},\cX{\ms_2},\dots)$ denotes the rate of
production/destruction of $\ms_j$ due to the reaction $\mr_i$, which
typically depends on the parameters $k_1,k_2,\dots$ that are intrinsic
to $\mr_i$ (reaction rate constants and stoichiometric coefficients)
and also on the concentrations of the reactants
$\cX{\ms_1},\cX{\ms_2}\cdots$ of $\mr_i$. The value of $\psi_{ji}$ is
either positive or negative depending on whether $\ms_j$ is produced
or consumed (respectively) by $\mr_i$, or zero if $\mr_i$ is not
involved in the production or consumption of $\ms_j$. 

\medskip

To facilitate the discussion, we use as a running example a simple
biological network consisting of species $\ms_1$, $\ms_2$, and
$\ms_3$, represented by the following set of chemical reactions:
\begin{equation}
  \label{eq:simpleeqn}
  \centering
  \begin{aligned}
    \mr_1:\orxna{\ms_1}{k_1\cX{\ms_1}}{\ms_2} \\
    \mr_2:\orxna{\ms_1}{\gamma_1\cX{\ms_1}}{\varnothing}
  \end{aligned}
  \qquad 
  \begin{aligned}
    \mr_3:\orxna{\ms_2}{k_2\cX{\ms_2}}{\ms_1}\\
    \mr_4:\orxna{\ms_2}{\gamma_2\cX{\ms_2}}{\varnothing}
  \end{aligned}
  \qquad
  \begin{aligned}
    \mr_5:\orxna{\ms_3}{k_3\cX{\ms_3}}{\ms_2},
  \end{aligned}
\end{equation}
which correspond to the following set of ODEs derived from MAK:
\begin{subequations}\label{eq:simple-odes}
  \begin{align}
    \dotcX{\ms_1} &= k_2\cX{\ms_2} - (\gamma_1 + k_1)\cX{\ms_1}\\
    \dotcX{\ms_2} &= k_1 \cX{\ms_1} + k_3 \cX{\ms_3} -(\gamma_2 + k_2)\cX{\ms_2}\\
    \dotcX{\ms_3} &= - k_3 \cX{\ms_3}.
  \end{align}
\end{subequations}
With respect to the general model \eqref{eq:MAK}, the $\psi_{ji}$ for $j
\in \{1,2,3\}$ and $i \in \{1,2,3,4,5\}$ are given by
\begin{equation*}
  \begin{aligned}
    \psi_{11}\big(k_1,\cX{\ms_1}\big) &= -k_1\cX{\ms_1}\\ 
    \psi_{12}\big(\gamma_1,\cX{\ms_1}\big) &= -\gamma_1\cX{\ms_1}\\
    \psi_{13}\big(k_2,\cX{\ms_2}\big) &= k_2\cX{\ms_2}
  \end{aligned}
  \qquad
  \begin{aligned}
    \psi_{21}\big(k_1,\cX{\ms_1}\big) &= k_1\cX{\ms_1}\\ 
    \psi_{23}\big(k_2,\cX{\ms_2}\big) &= -k_2\cX{\ms_2}\\
    \psi_{24}\big(\gamma_2,\cX{\ms_2}\big) &= -\gamma_2\cX{\ms_2}\\
    \psi_{25}\big(k_3, \cX{\ms_3}\big) &= k_3 \cX{\ms_3}
  \end{aligned}
  \qquad
  \begin{aligned}
    \psi_{35}\big(k_3,\cX{\ms_3}\big) = -k_3 \cX{\ms_3},
  \end{aligned}
\end{equation*}
and $0$ otherwise.

\subsubsection*{Directed Bipartite Species-Reactions (DBSR) graph}

When a biological network is large, writing down the system of
MAK ODEs is cumbersome and therefore much work has been done on
understanding the behavior of chemical reaction networks from a
graph-theoretic perspective \cite{domijan2008graph}. The Directed
Bipartite Species-Reaction (DBSR) graph representation of chemical
reaction networks was developed in \cite{ivanova1979conditions},
and is closely related to the Species-Reaction (SR) graph introduced
in \cite{craciun2006multiple}. In the construction of the DBSR graph,
every species in the network is assigned to an elliptical node, and every
chemical reaction is assigned to a rectangular node.
For every reaction in the network $\mr_i$, there exist directed edges
from the nodes corresponding to the reactants of $\mr_i$ to
the node $\mr_i$, and from the node
$\mr_i$ to the nodes corresponding to the products of $\mr_i$. It is
worth noting that this formulation is similar to that in
\cite{saez2005dissecting}, using storages and currents. The DBSR graph
of the network \eqref{eq:simpleeqn} is shown in
Figure~\ref{fig:dbgraphfull_simple}. From this graph, we can infer,
for example, that $\ms_2$ is produced from $\ms_1$ due to the reaction
$\mr_1$, and that $\ms_1$ is a reactant in $\mr_1$.  Therefore, the
concentration of $\ms_1$ is required in the computation of the rate of
the reaction $\mr_1$.
\begin{figure}[h]
  \centering
  \psfrag{S1}[][]{$\ms_1$}
  \psfrag{S2}[][]{$\ms_2$}
  \psfrag{S3}[][]{$\ms_3$}
  \psfrag{R1}[][]{$\mr_1$}
  \psfrag{R2}[][]{$\mr_2$}
  \psfrag{R3}[][]{$\mr_4$}
  \psfrag{R4}[][]{$\mr_3$}
  \psfrag{R5}[][]{$\mr_5$}
  \hfill\subfigure[\label{fig:dbgraphfull_simple}]{\includegraphics[width=.37\textwidth]{dbsr_simpleexample.ps}}
  \hfill
  \subfigure[\label{fig:ddbgraphfull_simple}]{\includegraphics[width=.37\textwidth]{ddbsr_simpleexample.ps}}\hspace*{\fill}
  \caption{\subref{fig:dbgraphfull_simple} DBSR graph and \subref{fig:ddbgraphfull_simple} Dynamic DBSR graph of the network
    represented by \eqref{eq:simpleeqn}}
\end{figure}

\subsubsection*{Dynamic DBSR graph}

While the DBSR graph is useful in understanding the overall structure
of a network of chemical reactions, it does not provide information
about the flow of information in the network. For instance, the graph
does not directly show whether the reaction $\mr_1$ affects the
dynamics of $\ms_1$. To obviate this problem, we define the
\emph{dynamic} DBSR graph, which is a DBSR graph overlaid with arrows
expressing the flow of information due to the dynamics of the
network. In this graph, a dashed arrow from a reaction node $\mr_i$ to
a product node $\ms_j$ indicates that $\cX{\ms_j}$ is affected by
$\mr_i$, usually by $\ms_j$ being consumed in the reaction. Just like
in the DBSR graph, a solid arrow from node $\mr_i$ to node $\ms_j$
indicates that $\ms_j$ is produced by the reaction, while a solid
arrow from node $\ms_j$ to node $\mr_i$ indicates that $\ms_j$ is a
reactant of $\mr_i$. The Dynamic DBSR graph of network
\eqref{eq:simpleeqn} is shown in Figure~\ref{fig:ddbgraphfull_simple}.


\subsection{Modular decomposition of a biological network}

The modular decomposition of a biological network represented by the MAK
ODE \eqref{eq:MAK} entails the assignment of each chemical species and
each chemical reaction in the network to modules. These modules then
need to be interconnected appropriately such that the ODE
\eqref{eq:MAK} can be reconstructed from the module dynamics. In
our framework, the assignment of a chemical species $\ms_j$ to a
module means that the species concentration $\cX{\ms_j}$ is part of
the state of that module alone. The assignment of a chemical reaction
to a module means that all the reaction parameters (such as
stoichiometric coefficients and rate constants) appear only in that
module. These observations lead to the formulation of two rules for a
modular decomposition:

\begin{rules}[Partition of species]\label{ru:species}
  Each chemical species $\ms_j$ must be associated with one and only one
  module $\mv$, and the state of $\mv$ is a vector containing the
  concentrations of all chemical species associated with $\mv$.\frqed
\end{rules}

\begin{rules}[Partition of reactions]\label{ru:reactions}
  Each chemical reaction $\mr$ must be associated with one and only
  one module $\mv$, and the stoichiometric parameters and rate
  constants associated with $\mr$ must only appear within the dynamics
  of module $\mv$.\frqed
\end{rules}

In terms of the DBSR graphs, Rules~\ref{ru:species}
and~\ref{ru:reactions} express that each node in the graph
(corresponding to either a species or a reaction) must be associated
with a single module. Therefore, our modular decomposition can be
viewed as a \emph{partition} of the nodes of the DBSR graphs. We
recall that a \emph{partition} of a graph is an assignment of the
nodes of the graph to disjoint sets.

\medskip

\begin{figure}[H]
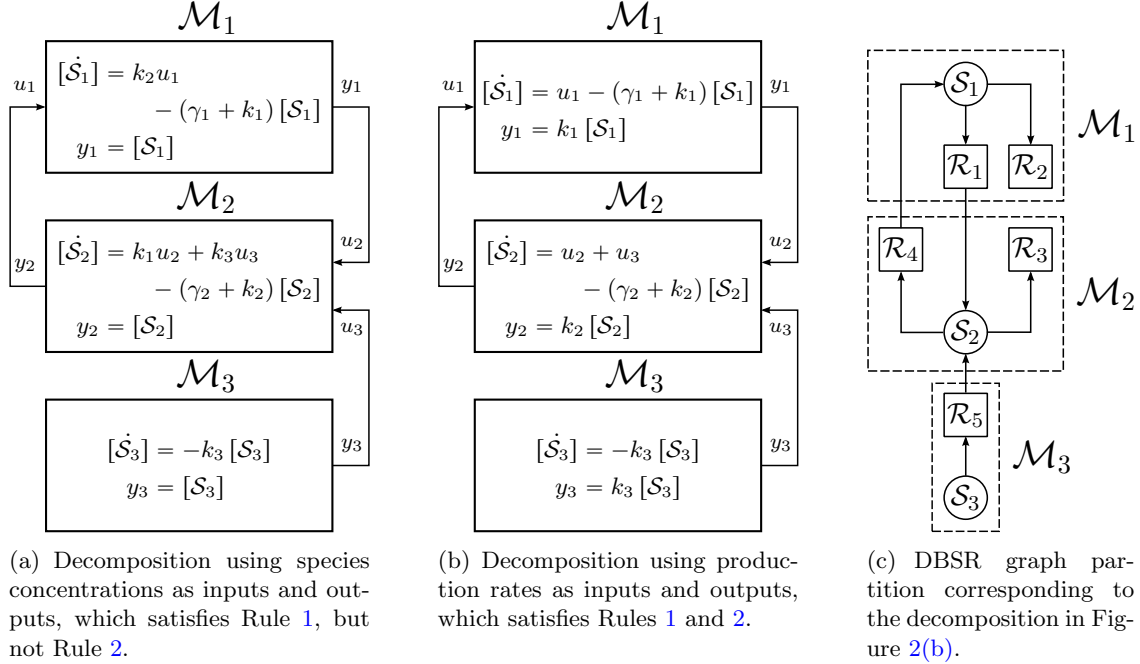

  \centering 
  \psfrag{S1}[][]{$\ms_1$}
  \psfrag{S2}[][]{$\ms_2$}
  \psfrag{S3}[][]{$\ms_3$}
  \psfrag{R1}[][]{$\mr_1$}
  \psfrag{R2}[][]{$\mr_2$}
  \psfrag{R3}[][]{$\mr_3$}
  \psfrag{R4}[][]{$\mr_4$}
  \psfrag{R5}[][]{$\mr_5$}
  \psfrag{M1}[l][l]{\Large$\mv_1$}
  \psfrag{M2}[l][l]{\Large$\mv_2$}
  \psfrag{M3}[l][l]{\Large$\mv_3$}
  \psfrag{u1}[][]{\scriptsize$u_1$}
  \psfrag{u2}[][]{\scriptsize$u_2$}
  \psfrag{u3}[][]{\scriptsize$u_3$}
  \psfrag{y1}[][]{\scriptsize$y_1$}
  \psfrag{y2}[][]{\scriptsize$y_2$}
  \psfrag{y3}[][]{\scriptsize$y_3$}
  \psfrag{a}[][]{\parbox{3in}{\footnotesize\begin{align*}\dotcX{\ms_1} &= k_2 u_1\\ &\qquad - (\gamma_1 + k_1)\cX{\ms_1}\\ y_1 &= \cX{\ms_1}\end{align*}}}
  \psfrag{b}[][]{\parbox{3in}{\footnotesize\begin{align*}\dotcX{\ms_2} &= k_1 u_2 + k_3 u_3\\ &\qquad - (\gamma_2 + k_2)\cX{\ms_2}\\ y_2 &= \cX{\ms_2}\end{align*}}}
  \psfrag{c}[][]{\parbox{3in}{\footnotesize\begin{align*}\dotcX{\ms_3} &= -k_3\cX{\ms_3}\\ y_3 &= \cX{\ms_3}\end{align*}}}
 \hfill  \subfigure[Decomposition using species concentrations as inputs and
  outputs, which satisfies Rule~\ref{ru:species}, but not Rule~\ref{ru:reactions}.\label{fig:simpleexample}]{\includegraphics[width=.3\textwidth]{blockexample.ps}}
  \psfrag{a}[][]{\parbox{3in}{\footnotesize\begin{align*}\dotcX{\ms_1} &= u_1 - (\gamma_1 + k_1)\cX{\ms_1}\\ y_1 &= k_1\cX{\ms_1}\end{align*}}}
   \psfrag{b}[][]{\parbox{3in}{\footnotesize\begin{align*}\dotcX{\ms_2} &= u_2 + u_3\\ &\qquad - (\gamma_2 + k_2)\cX{\ms_2}\\ y_2 &= k_2\cX{\ms_2}\end{align*}}}
   \psfrag{c}[][]{\parbox{3in}{\footnotesize\begin{align*}\dotcX{\ms_3} &= -k_3\cX{\ms_3}\\ y_3 &= k_3\cX{\ms_3}\end{align*}}}
  \hfill \subfigure[Decomposition using production rates as inputs and
  outputs, which satisfies Rules~\ref{ru:species} and~\ref{ru:reactions}.\label{fig:simpleexample_correct}]{\includegraphics[width=.3\textwidth]{blockexample.ps}}
\hfill \subfigure[DBSR
  graph partition corresponding to the decomposition in Figure~\ref{fig:simpleexample_correct}. \label{fig:dbgraphpartition2_simple}]{\includegraphics[width=0.2\columnwidth]{dbsr_simpleexample_partition.ps}}\hspace*{\fill}
  \caption{Modular decomposition of the biochemical network
    \eqref{eq:simpleeqn} corresponding to the dynamic DBSR graph in
    Figure~\ref{fig:ddbgraphfull_simple}.}
  \label{fig:2_partition}
\end{figure}

The choice of signals used to communicate between modules has a direct
impact on whether or not Rules~\ref{ru:species} and~\ref{ru:reactions}
are violated. To understand why this is so, consider again the
biological network \eqref{eq:simpleeqn}, corresponding to the MAK ODEs
given by \eqref{eq:simple-odes}, and suppose that we want to associate
each of the three species $\ms_1$, $\ms_2$ and $\ms_3$ with a
different component.  Figure~\ref{fig:2_partition} shows two
alternative partitions of the network that satisfy
Rule~\ref{ru:species} and exhibit the dynamic modularity
property. That is, the concentration of each species appears in the
state of one and only one component, and when the three components are
combined, we obtain precisely the MAK ODEs in \eqref{eq:simple-odes}.

\medskip

In the decomposition depicted in the block diagram in
Figure~\ref{fig:simpleexample}, the communicating signals are the
\emph{concentrations} of the species. Specifically, the outputs $y_1$,
$y_2$, and $y_3$ of components $\mv_1$, $\mv_2$, and $\mv_3$,
respectively, are the concentrations of the species $\ms_1$, $\ms_2$,
and $\ms_3$; which in turn are the inputs $u_2$, $u_1$, or $u_3$ of
components $\mv_2$, $\mv_1$, and $\mv_3$, respectively.  This type of
decomposition, where
species concentrations are used as communicating signals between
modules, has been commonly done in the literature
\cite{papachris,sontag2011modularity,saez2005dissecting,saez2008automatic,saez2004modular}.
However, this decomposition violates Rule~\ref{ru:reactions}, because
the rate parameters of the reactions $\mr_1$, $\mr_3$, and $\mr_5$
appear in multiple components. Consequently, this partition does not
exhibit parametric modularity; for example, a change in the rate
constant $k_3$ for reaction $\mr_5$ would change the internal dynamics
of modules $\mv_2$ and $\mv_3$. We thus do not refer to the components
$\mv_1$, $\mv_2$, and $\mv_3$ in Figure~\ref{fig:simpleexample} as
``modules.''

\medskip

An alternative choice for the communicating signals would be the
\emph{rates of production} of the species, which leads to the
decomposition depicted in the block diagram in
Figure~\ref{fig:simpleexample_correct}. With this decomposition, we
can now partition both the species and the reaction nodes among the
different components so that the parameters of each reaction are
confined to a single module, as illustrated in
Figure~\ref{fig:dbgraphpartition2_simple}. We thus have a modular
decomposition that simultaneously satisfies Rules~\ref{ru:species}
and~\ref{ru:reactions} and refer to the components $\mv_1$, $\mv_2$,
and $\mv_3$ in Figure~\ref{fig:simpleexample_correct} as
``modules.''

\subsection{Rates as communicating signals}
\label{subsec:rates}

In the context of a simple example, we have seen that using
\emph{rates} as the communicating signals between modules (as opposed
to protein concentrations) enables a modular decomposition that
simultaneously satisfies Rules~\ref{ru:species} and~\ref{ru:reactions}. We now generalize these ideas to arbitrary biological
networks.

\medskip

When partitioning a dynamic DBSR graph into modules, each arrow of the
graph ``severed'' by the partition corresponds to an interconnecting
signal flowing between the resulting modules, in the direction of the
arrow. For example, one can see two arrows being severed in Figure~\ref{fig:dbgraphpartition2_simple} by the partition between modules
$\mv_1$ and $\mv_2$ and then two signals ($y_2$ and $y_1$) connecting
the corresponding modules in Figure~\ref{fig:simpleexample_correct}. In the remainder of this section, we
present two basic scenarios that can arise in partitioning a network
into two modules and discuss the signals that must flow between these
modules. These two cases can be applied iteratively to partition a
general network into an arbitrary number of modules.

\subsubsection*{Partition at the output of a reaction node.}
\begin{figure}[h]
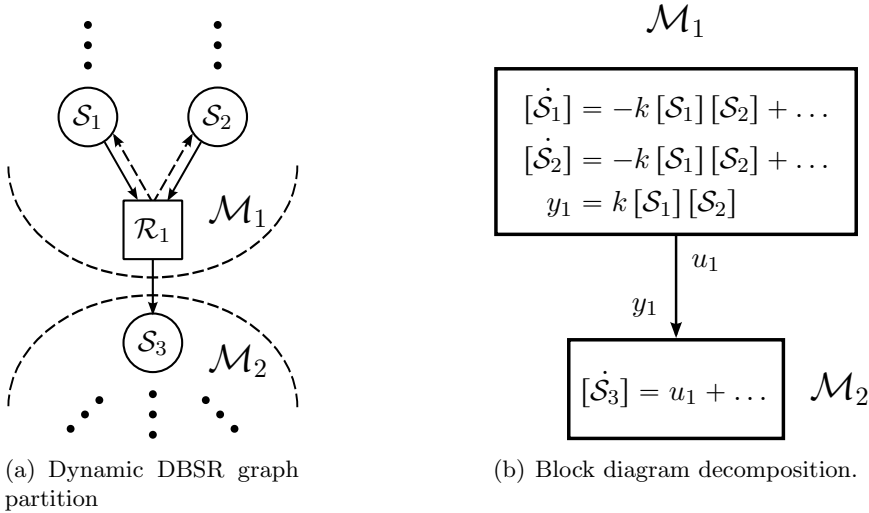

  \hfill
  \psfrag{S1}[][]{$\ms_1$}
  \psfrag{S2}[][]{$\ms_2$}
  \psfrag{S3}[][]{$\ms_3$}
  \psfrag{R1}[][]{$\mr_1$}
  \psfrag{M1}[][]{\Large$\mv_1$}
  \psfrag{M2}[][]{\Large$\mv_2$}
  \psfrag{u1}[][]{$u_1$}
  \psfrag{y1}[][]{$y_1$}
  \psfrag{a}[][]{\parbox{3in}{\begin{align*}
  			\dotcX{\ms_1} &= -k\cX{\ms_1}\cX{\ms_2} + \dots\\
  			\dotcX{\ms_2} &= -k\cX{\ms_1}\cX{\ms_2} + \dots\\
  			y_1 &= k\cX{\ms_1}\cX{\ms_2}
  		\end{align*}}}
  \psfrag{b}[][]{$\dotcX{\ms_3} = u_1 + \dots$}
  \subfigure[Dynamic DBSR graph partition\label{fig:rateexample-1}]{\includegraphics[width=.27\textwidth]{dbsrpartition1.ps}}
  \hfill
  \subfigure[Block diagram decomposition. \label{fig:rateexample-2}]{\includegraphics[width=.33\textwidth]{blockpartition1.ps}}
  \hspace*{\fill}
  \caption{Modular decomposition corresponding to the partition of a
    biochemical network at the output of a reaction node of the
    dynamic DBSR graph.}
\end{figure}
Suppose first that we partition a biological network into two modules
$\mv_1$ and $\mv_2$ at the \emph{output of a reaction node} of the
dynamic DBSR graph, corresponding to a generic elementary reaction of
the form

\begin{equation}
\label{eq:generic}
\mr_1:  \trxn{\ms_1}{\ms_2}{k \cX{\ms_1}\cX{\ms_2}}{\ms_3},
\end{equation}
as shown in Figure~\ref{fig:rateexample-1}. This can be accomplished
by connecting an output $y_1$ from module $\mv_1$ to an input $u_2$ of
module $\mv_2$ that is equal to the rate of production of $\ms_3$ due
to $\mr_1$, which is given by $k \cX{\ms_1}\cX{\ms_2}$. The block
diagram representation of this partition is shown in
Figure~\ref{fig:rateexample-2}. In this configuration, the reaction
rate parameter $k$ only appears inside the module $\mv_1$ and we thus
have parametric modularity. As far as $\mv_2$ is concerned, the rate
of production of $\ms_3$ in \emph{molecules per unit time} is given by
the abstract chemical reaction
\begin{equation*}
 \zrxn{u_1}{\ms_3},
\end{equation*}
where the rate $u_1$ is an input to the module. In this decomposition,
we also have dynamic modularity, since when we combine the dynamics of
the two modules in Figure~\ref{fig:rateexample-2}, we recover the MAK
ODEs. This partition therefore ensures that both
Rules~\ref{ru:species} and~\ref{ru:reactions} are satisfied. We
emphasize that the \emph{single arrow} from node $\mr_1$ to node
$\ms_3$ that is ``severed'' by the partition in
Figure~\ref{fig:rateexample-1}, gives rise to \emph{one signal} flowing from
$\mv_1$ to $\mv_2$ in Figure~\ref{fig:rateexample-2}.

\begin{figure}[h]
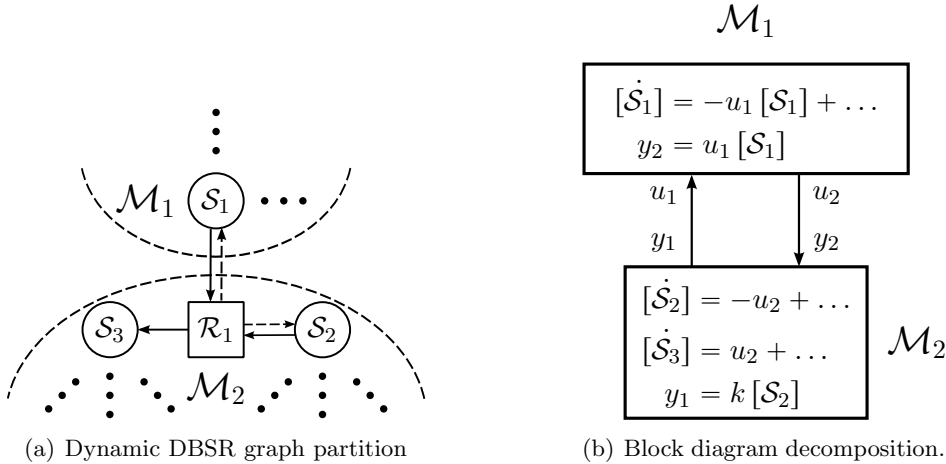

  \hfill
    \psfrag{S1}[][]{$\ms_1$}
    \psfrag{S2}[][]{$\ms_2$}
    \psfrag{S3}[][]{$\ms_3$}
    \psfrag{R1}[][]{$\mr_1$}
    \psfrag{M1}[][]{\Large$\mv_1$}
    \psfrag{M2}[][]{\Large$\mv_2$}
    \psfrag{u1}[][]{$u_1$}
    \psfrag{y1}[][]{$y_1$}
    \psfrag{u2}[][]{$u_2$}
    \psfrag{y2}[][]{$y_2$}
    \psfrag{a}[][]{\parbox{3in}{\begin{align*}
    		\dotcX{\ms_1} &= -u_1\cX{\ms_1} + \dots\\
    		y_2 &= u_1\cX{\ms_1}
    		\end{align*}}}
    \psfrag{b}[][]{\parbox{3in}{\begin{align*}
    		\dotcX{\ms_2} &= -u_2 + \dots\\
    		\dotcX{\ms_3} &= u_2 + \dots\\
    		y_1 &= k\cX{\ms_2}
    		\end{align*}}}
  \subfigure[Dynamic DBSR graph partition\label{fig:dbso1}]{\includegraphics[width=.35\textwidth]{dbsrpartition2.ps}}
  \hfill
  \subfigure[Block diagram decomposition. \label{fig:dbso2}]{\includegraphics[width=.3\textwidth]{blockpartition2.ps}}
  \hspace*{\fill}
  \caption{Modular decomposition corresponding to the partition of a
    biochemical network at the output of a species node of the dynamic
    DBSR graph. }
\label{fig:dbgraphsecondorder1}
\end{figure}

\subsubsection*{Partition at the input of a reaction node.}
Now consider the case where we partition the network at an
\emph{input of a reaction node} of the dynamic DBSR graph,
corresponding to a generic elementary reaction of the form in
\eqref{eq:generic}, as shown in Figure~\ref{fig:dbso1}. This can be
accomplished by a \emph{bidirectional} connection between the two
modules: The output $y_1$ from $\mv_2$ is connected to the input $u_1$
of $\mv_1$ and is equal to $k \cX{\ms_2}$, which is the
\emph{degradation} rate of a single molecule of $\ms_1$ due to
the reaction $\mr_1$, in \emph{molecules per molecule of $\ms_1$ per unit
  time}. The output $y_2$ from $\mv_1$ is connected to the input $u_2$
of $\mv_2$ and is equal to $u_1\cX{\ms_1}$, which is the rate of
production of $\ms_3$ and the net consumption rate of $\ms_2$ due to
$\mr_1$, both in \emph{molecules per unit time}. The block diagram
representation of this modular decomposition is shown in Figure~\ref{fig:dbso2}. In isolation, the block $\mv_1$ corresponds to a
chemical reaction of the form
\begin{align*}
  \orxn{\ms_1}{u_1 \cX{\ms_1}}{\varnothing},
\end{align*}
where the input $u_1$ determines the degradation rate of the species
$\ms_1$, and the block $\mv_2$ corresponds to an abstract chemical
reaction of the form
\begin{align*}
  \orxn{\ms_2}{u_2}{\ms_3},
\end{align*}
where the input $u_2$ determines the production rate of the species
$\ms_3$, which is also the net consumption rate of $\ms_2$. This
decomposition is parametrically modular since the reaction rate
parameter $k$ is only part of the module $\ms_2$, which contains the
reaction $\mr_1$.
We emphasize that the \emph{two arrows} between $\mr_1$ to $\ms_1$
that are ``severed'' by the partition in the dynamic DBSR in
Figure~\ref{fig:dbso1} give rise to the \emph{two signals} flowing
between $\mv_1$ and $\mv_2$ in Figure~\ref{fig:dbso2}.

\medskip

The discussion above gives rise to the following general rule that
governs the communicating signals between modules.

\begin{rules}[Communication signals between modules]
  Each arrow of the dynamic DBSR graph that is ``severed'' by the
  partition that defines the modular decomposition gives rise to one
  signal that must flow between the corresponding
  modules. Specifically,
  \label{ru:dbg}
  \begin{enumerate}
  \item When the modular decomposition cuts the dynamic DBSR graph
    between a reaction node $\mr_i$ and a product species node $\ms_j$ at the output of node $\mr_i$,
    one signal must flow between the modules: the module with the
    reaction must have an output equal to the rate [in molecules per unit time] at which the product $\ms_j$ is produced by the
    reaction.

  \item When the modular decomposition cuts the dynamic DBSR graph
    between a reaction node $\mr_i$ and a reactant species node $\ms_j$ at the output of node $\ms_j$, two signals must flow between the corresponding modules:
    the module with the reaction must have an output equal to the rate
    at which each molecule of $\ms_j$ is degraded [in molecules per
    molecule of $\ms_j$ per unit time], and the module with $\ms_j$
    must have an output equal to the total rate at which the molecules
    of $\ms_j$ are consumed [in molecules per unit time].\frqed
\end{enumerate}
\end{rules}

When Rules~\ref{ru:species}--\ref{ru:dbg} are followed, the
decomposition of the network into modules will exhibit both parametric and
dynamic modularity. Furthermore, the network can be decomposed into
any number of modules less than or equal to the total number of
reacting species in the network.

\begin{remark}
  In exploring the different possible cases, we restricted our
  discussion to elementary reactions (at most two reactants) and
  assumed that at least one of the reactants remains in the same
  module as the reaction. We made these assumptions mostly for
  simplicity, as otherwise one would have to consider a large
  number of cases.\frqed
\end{remark}


\subsection{Summation Junctions}

In biological networks, it is not uncommon for a particular species to
be produced or degraded by two or more distinct pathways. In fact, we
have already encountered this in the biochemical network
\eqref{eq:simpleeqn} corresponding to the dynamic DBSR graph depicted
in Figure~\ref{fig:ddbgraphfull_simple}, where the species $\ms_2$ is
produced both by reactions $\mr_1$ and $\mr_5$. The use of rates as
communicating signals between modules allows for the use of summation
junctions outside modules to combine different mechanisms to
produce/degrade a chemical species.
\begin{figure}[h]
  \centering
  \psfrag{M1}[l][l]{\Large$\mv_1$}
  \psfrag{M2}[l][l]{\Large$\mv_2$}
  \psfrag{M3}[l][l]{\Large$\mv_3$}
  \psfrag{u1}[][]{\scriptsize$u_1$}
  \psfrag{v}[][]{\scriptsize$v$}
  \psfrag{y1}[][]{\scriptsize$y_1$}
  \psfrag{y2}[][]{\scriptsize$y_2$}
  \psfrag{y3}[][]{\scriptsize$y_3$}
   \psfrag{a}[][]{\parbox{3in}{\footnotesize\begin{align*}\dotcX{\ms_1} &= u_1 - (\gamma_1 + k_1)\cX{\ms_1}\\ y_1 &= k_1\cX{\ms_1}\end{align*}}}
   \psfrag{b}[][]{\parbox{3in}{\footnotesize\begin{align*}\dotcX{\ms_2} &= v - (\gamma_2 + k_2)\cX{\ms_2}\\ y_2 &= k_2\cX{\ms_2}\end{align*}}}
   \psfrag{c}[][]{\parbox{3in}{\footnotesize\begin{align*}\dotcX{\ms_3} &= -k_3\cX{\ms_3}\\ y_3 &= k_3\cX{\ms_3}\end{align*}}}
\includegraphics[width=0.3\textwidth]{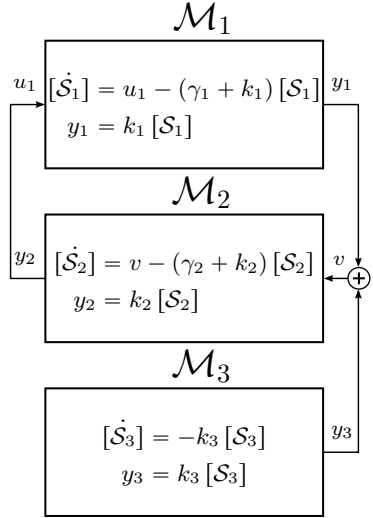}
\caption{Modular decomposition from Figure \ref{fig:simpleexample_correct}, simplified using summation junctions.}
  \label{fig:simpleexample_correctsummed}
\end{figure}
This is illustrated in Figure~\ref{fig:simpleexample_correctsummed},
where we provide a modular decomposition alternative to that shown in
Figure~\ref{fig:simpleexample_correct}. This decomposition still
preserves the properties of dynamic and parametric modularity, but
permits simpler blocks than those in
Figure~\ref{fig:simpleexample_correct}, since each module now only has
a single input and a single output (SISO).  As we shall see in
Section~\ref{sec:negfdbk}, there are many tools that can be used to
analyse interconnections of SISO modules that include external
summation junctions.

\medskip

It is worth noting that when species concentrations are used as the
communicating signals between modules [as in
Figure~\ref{fig:simpleexample}], it is generally not possible to use
summation junctions to combine two or more distinct mechanisms to
produce or degrade a chemical species. Even if parametric modularity
were not an issue, this limitation would typically lead to more
complicated modules with a larger number of inputs and outputs.


\section{Common Modules}
\label{sec:examples}

In this section, we consider a few key biological modules that arise
in gene regulatory networks, enzymatic networks, and signaling
pathways. We characterize these modules in terms of system theoretic
properties that can be used to establish properties of complex
interconnections involving these modules. Before introducing the
biological modules of interest, we briefly recall some of these system
theoretic properties.

\subsection{Module properties}
\label{subsec:modprop}
Consider a generic input-output module, expressed by an ODE of the
form
\begin{align}\label{eq:nonlinear}
  \dot x&=A(x,u), &
  y&= B(x,u), & 
  x\in\R^n,u\in\R^k,y\in\R^m,
\end{align}
where $x(t)$ denotes the $n$-vector state of the module, $u(t)$ the
$k$-vector input to the module, and $y(t)$ the $m$-vector output from
the module.

\medskip

We say that the module described by \eqref{eq:nonlinear} is
\emph{positive} if the entries of its state vector $x(t)$ and output
vector $y(t)$ never take negative values, as long as all the entries
of the initial condition $x(0)$ and of the input vector $u(t)$,
$\forall t \ge 0$ never take negative values. All the modules
described in this section are positive.

\medskip

We say that the module described by \eqref{eq:nonlinear} is
\emph{cooperative} (also known as \emph{monotone with respect to the
  positive orthant}) if for all initial conditions $x_0,\bar
x_0\in\R^n$ and inputs $u(t), \bar u(t)\in\R^k$, $\forall t\ge 0$, we
have that
\begin{align*}
  x_0\gg \bar x_0 \quad&\&\quad u(t)\succeq \bar u(t),\;\forall t\ge 0 
  \imply 
  x(t;x_0,u)\gg x(t;\bar x_0,\bar u) ,\;\forall t> 0
\end{align*}
where $x(t;x_0,u)$ denotes the solution to \eqref{eq:nonlinear} at
time $t$, starting from the initial condition $x(0)=x_0$ and with the
input $u$. Given two vectors $v,\bar v$, we write $v\gg \bar v$ if
every entry of $v$ is strictly larger than the corresponding entry of
$\bar v$ and we write $v\succeq \bar v$ if every entry of $v$ is
larger than or equal to the corresponding entry of $\bar v$. The reader is referred to \cite{angeli2003monotone,angeli2004interconnections,angeli2004multi}
for a more comprehensive treatment of monotone dynamical systems,
including simple conditions to test for monotonicity and results that
allow one to infer monotonicity of a complex network from the
monotonicity of its constituent parts. Several modules described in
this section are cooperative.

\medskip

The \emph{Input-to-State Static Characteristic Function (ISSCF)
  $g(u^*)$ of \eqref{eq:nonlinear}} specifies how a constant input
$u(t)=u^*$, $\forall t\ge 0$ to the module maps to the corresponding
equilibrium value of the state $x(t)=x^*$, $\forall t\ge 0 $. In terms
of \eqref{eq:nonlinear}, the value of $g(u^*)$ is the (unique)
solution $x^*$ to the steady-state equation $A(x^*,u^*)$ = 0. When
this equation has multiple solutions $x^*$, the ISSCF is not well
defined.

For modules with a well-defined ISSCF, the \emph{Input-to-Output
  Static Characteristic Function (IOSCF) $f(u^*)$ of
  \eqref{eq:nonlinear}} specifies how a constant input $u(t)=u^*$,
$\forall t\ge 0$ to the module maps to the corresponding equilibrium
value of the output $y(t)=y^*$, $\forall t\ge 0$. In terms of
\eqref{eq:nonlinear}, the value of $f(u^*)$ is given by
$B(g(u^*),u^*)$. We shall see in Section~\ref{sec:negfdbk} that one can determine the equilibrium point of a network obtained from the interconnection of
several input-output modules like \eqref{eq:nonlinear}, from the
IOSCFs and the ISSCFs of the constituent modules.

\medskip

For systems with a well-defined ISSCF, the \emph{Linearized Transfer
  Function (LTF) $H(s)$ of \eqref{eq:nonlinear} around an equilibrium
  defined by the input $u^*$} determines how a small perturbation
$\delta u(t)\eqdef u(t)-u^*$ of the input $u(t)$ around the constant
input $u(t)=u^*$, $\forall t\ge 0$ leads to a perturbation $\delta
y(t)\eqdef y(t)-y^*$ of the output $y(t)$ around the constant
equilibrium output $y(t)= y^*\eqdef f(u^*)$, $\forall t\ge 0$. In
particular, $\delta y(t) =\scr{L}^{-1}[H(s)] \star \delta u(t)$, where
$\scr{L}^{-1}[H(s)]$ denotes the inverse Laplace transform of $H(s)$
and $\star$ the convolution operator \cite{Hespanha09}. We shall also
see in Section~\ref{sec:negfdbk} that one can determine the LTF of a
network obtained from the interconnection of several input-output
modules like \eqref{eq:nonlinear}, from the LTFs of the constituent
modules.

\medskip

The LTF of a module like \eqref{eq:nonlinear} is given by a rational
function and generally, the (local) stability of the equilibrium
defined by the input $u^*$ can be inferred from the roots of the
denominator of the LTF. Specifically, if all the roots have strictly
negative real parts, in which case we say that the LTF is
\emph{bounded-input/bounded-output (BIBO) stable}, then the
equilibrium point is locally asymptotically stable, which means
solutions starting close to the equilibrium will converge to it as $t
\to \infty$; this assumes that the McMillan degree of the LTF equals
the size $n$ of the state $x$ of \eqref{eq:nonlinear}
\cite{Hespanha09}, which is generically true, but should be tested.

\subsection{Transcriptional regulation (TR) module}
\label{subsec:transreg}

A gene regulatory network consists of a collection of transcription
factor proteins, each involved in the regulation of other proteins in
the network. Such a network can be decomposed into
\emph{transcriptional regulation (TR) modules}, each containing a
transcription factor $\ms_0$, the promoter regions of a set of genes
$\gene{1},\gene{2},\dots,\gene{F}$ that $\ms_0$ up-regulates or
down-regulates, and the corresponding mRNA molecules $mRNA_1,mRNA_2,\dots,mRNA_F$ transcribed. The case $F>1$ is referred to in the literature as
\emph{fan-out} \cite{kim2010fan}.

\medskip

The input $u$ to a TR module is the rate of production of $\ms_0$ due to
exogenous processes such as regulation from other TR modules, and can be associated
with a generic reaction of the form
\begin{align*}
  \zrxn{u}{\ms_0}.
\end{align*}
A number $q_j\ge 1$ of molecules of the transcription factor $\ms_0$
can bind to the promoter region $P_j$ of the gene $\gene{j}$, which is
represented by the reaction
\begin{align*}
  q_j\ms_0 + P_j &\xrightleftharpoons[k^\mrm{off}_j\cX{\bndxy{\ms_0}{P_j}}]{k^\mrm{on}_j\cX{\ms_0}^{q_j}\cX{P_j}}\bndxy{\ms_0}{P_j}, 
  \qquad \forall j\in \{1,2,\dots,m\}.
\end{align*}

The total concentration of promoter regions
$P^\mrm{tot}_j=[P_j]+[\bndxy{\ms_0}{P_j}]$ for the gene $\gene{j}$
(bound and unbound to the transcription factor) is assumed to remain
constant. When $\ms_0$ \emph{activates} the gene $\gene{j}$, the bound
complex $\bndxy{\ms_0}{P_j}$ gives rise to transcription, which is
expressed by a reaction of the form
\begin{align*}
  \orxna{\bndxy{\ms_0}{P_j}}{\alpha_j\cX{\bndxy{\ms_0}{P_j}}}{\bndxy{\ms_0}{P_j} + mRNA_j}.
\end{align*}
Alternatively, when $\ms_0$ \emph{represses} the gene $\gene{j}$, it is the
unbounded promoter $P_j$ that gives rise to transcription, which is
expressed by a reaction of the form
\begin{align*}
  \orxna{P_j}{\alpha_j\cX{P_j}}{P_j + mRNA_j}.
\end{align*}
Additional reactions in the module include the translation of $mRNA_j$ to $\ms_j$
\begin{align*}
\orxna{mRNA_j}{\beta_j\cX{mRNA_j}}{mRNA_j + \ms_j},
\end{align*}
and the protein and mRNA degradation reactions
\begin{align*}
\orxna{\ms_0}{\bar{\beta}\cX{\ms_0}}{\varnothing} ,\\
\orxna{mRNA_j}{\gamma_j\cX{mRNA_j}}{\varnothing}
\end{align*}
The TR module has $F$ outputs $y_1,y_2,\dots, y_F$ that are equal to the rates of
translations of the proteins $\ms_{1},\ms_{2},\dots,\ms_{F}$, respectively. In particular,
\begin{align*}
y_j = \beta_j\cX{mRNA_j}
\end{align*}
When $F=1$, we refer to each module simply as a TR activator or TR repressor module, and the subscript $j$'s can be omitted. 

Table~\ref{tab:mod} shows the system of ODEs that correspond to the TR module, as well as its IOSCF and LTF, under the following assumptions:
\begin{assumption}[Homogeneity in TR module]
  \label{as:tr-homogeneous}
  For simplicity of presentation, it is assumed that the association
  and dissociation constants, the total promoter concentration, and
  the stoichiometric coefficients are the same for every gene,
  i.e., that $k^\mrm{off}_j=k^\mrm{off}$, $k^\mrm{on}_j=k^\mrm{on}$,
  $\ptot_j = \ptot$ and $q_j = q$, $\forall
  j\in\{1,2,\dots,F\}$. 
\end{assumption}
\begin{assumption}[Parameters in TR module]
  \label{as:tr-timescale}
  The following assumptions on the parameter values are considered:
  \begin{enumerate}
  \item \label{as:time-scales} The binding-unbinding reactions are on
    timescales much faster than those of the transcription, translation,
    and decay reactions, i.e., $\koff,\kon \gg \gamma$, $\beta_j$,
    $\bbeta$ and $u(t)$ $\forall t\geq 0$ \cite{DelVecchio08}. This
    assumption simplifies the LTF of the module, as shown in Table~\ref{tab:mod}.
  \item \label{as:dissociation} The dissociation constant $K =
    \frac{\koff}{\kon}$ is much higher than the total promoter
    concentration i.e $K\gg\ptot$, implying that the affinity of each
    binding site is small.
    \frqed
  \end{enumerate}
\end{assumption}
An interesting consequence of Assumption~\ref{as:tr-timescale} is
that the LTFs from a perturbation in the input to a perturbation in each of the outputs do not depend
on the fan-out, which is not true for the original dynamics without
this assumption. For completeness, we include in Appendix~\ref{app:modchar} the LTF of the TR module computed without Assumption~\ref{as:tr-timescale}.

\medskip

\begin{figure}[h]
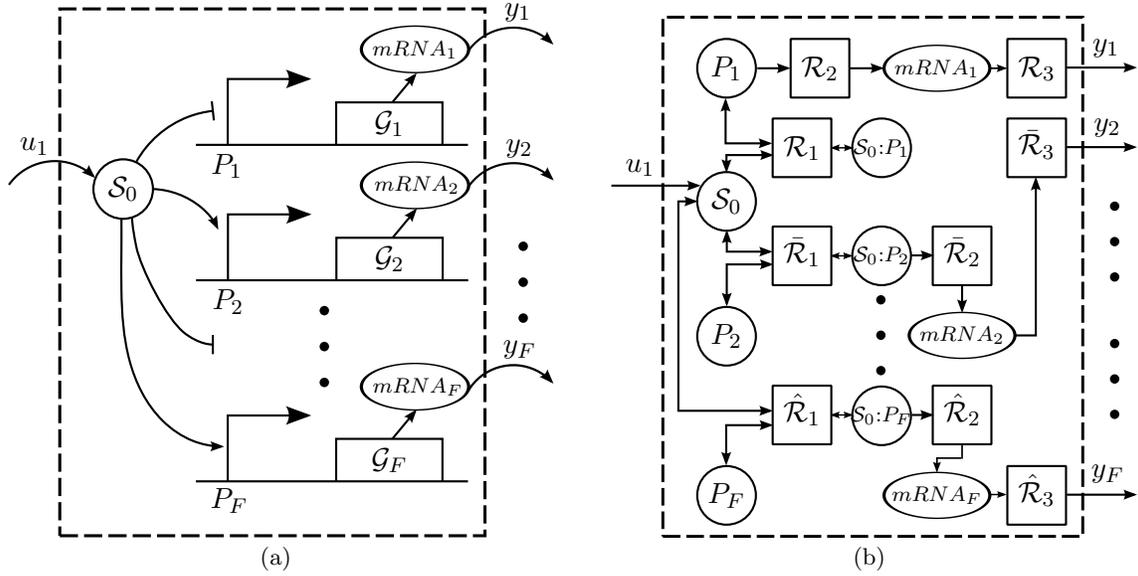

	\centering 
	{\scriptsize
		\psfrag{m1}[][]{$mRNA_1$}
		\psfrag{m2}[][]{$mRNA_2$}
		\psfrag{mF}[][]{$mRNA_F$}
		\psfrag{S0}[][]{\normalsize$\ms_0$}
		\psfrag{R1}[][]{\normalsize$\mr_1$}
		\psfrag{R2}[][]{\normalsize$\mr_2$}
		\psfrag{R3}[][]{\normalsize$\mr_3$}
		\psfrag{B1}[b][b]{\normalsize$\bar\mr_1$}
		\psfrag{B2}[b][b]{\normalsize$\bar\mr_2$}
		\psfrag{B3}[b][b]{\normalsize$\bar\mr_3$}
		\psfrag{H1}[b][b]{\normalsize$\hat\mr_1$}
		\psfrag{H2}[b][b]{\normalsize$\hat\mr_2$}
		\psfrag{H3}[b][b]{\normalsize$\hat\mr_3$}
		\psfrag{P1}[][]{\normalsize$P_1$}
		\psfrag{P2}[][]{\normalsize$P_2$}
		\psfrag{PF}[][]{\normalsize$P_F$}
		\psfrag{C1}[][]{$\bndxy{\ms_0}{P_1}$}
		\psfrag{C2}[][]{$\bndxy{\ms_0}{P_2}$}
		\psfrag{CF}[][]{$\bndxy{\ms_0}{P_F}$}
		\psfrag{G1}[][]{\normalsize$\gene{1}$}
		\psfrag{G2}[][]{\normalsize$\gene{2}$}
		\psfrag{GF}[][]{\normalsize$\gene{F}$}
		\psfrag{u1}[][]{\normalsize$u_1$}
		\psfrag{y1}[][]{\normalsize$y_1$}
		\psfrag{y2}[][]{\normalsize$y_2$}
		\psfrag{yF}[][]{\normalsize$y_F$}
		\hfill\subfigure[\label{fig:biotr}]{\includegraphics[width = 0.44\textwidth]{biotr_fanout.ps}}\hfill
		\subfigure[\label{fig:dbsrtr}]{\includegraphics[width = 0.44\textwidth]{dbsrtr_fanout.ps}}\hspace*{\fill}
		\caption{\subref{fig:biotr} Biological representation and \subref{fig:dbsrtr} DBSR graph of a TR module. For simplicity, protein degradation reactions are shown neither in Figure \ref{fig:dbsrtr} nor in all subsequent DBSR graphs presented in this paper.}
		\label{fig:tr}
	}
\end{figure}

\begin{table}[h]
\centering
\scalebox{0.9}{
\begin{tabular}{|>{\centering\arraybackslash}m{15ex}|>{\centering\arraybackslash}m{.4\textwidth}|>{\centering\arraybackslash}m{.46\textwidth}|}
  \hline
    &  $\ms_0$ activates $\gene{j}$
    &  $\ms_0$ represses $\gene{j}$ \\ \hline
  Dynamics
  (under Assumption~\ref{as:tr-homogeneous})&
  \multicolumn{2}{|c|}{\parbox{.8\textwidth}{
      \begin{align*}
        \dotcX{\ms_0} &= u -\bar{\beta} \cX{\ms_0} +  \sum_{j=1}^m q\big(
        k^\mrm{off} \cX{\bndxy{\ms_0}{P_j}} 
        - k^\mrm{on}\cX{\ms_0}^q(P^\mrm{tot}-\cX{\bndxy{\ms_0}{P_j}})\big)\\
        \dotcX{\bndxy{\ms_0}{P_j}} &= -k^\mrm{off} \cX{\bndxy{\ms_0}{P_j}} 
        + k^\mrm{on}\cX{\ms_0}^q(P^\mrm{tot}-\cX{\bndxy{\ms_0}{P_j}})
      \end{align*}
    }}\\ \cline{2-3}
    & $\dotcX{mRNA_j} = \alpha_j\cX{\bndxy{\ms_0}{P_j}}  - \gamma_j \cX{mRNA_j}$ & $\dotcX{mRNA_j} = \alpha_j(\ptot - \cX{\bndxy{\ms_0}{P_j}} )  - \gamma_j \cX{mRNA_j}$ \\ \hline
  Dynamics
  (under Assumptions~\ref{as:tr-homogeneous}--\ref{as:tr-timescale})&\multicolumn{2}{|c|}{$\displaystyle \dotcX{\ms_0} = u -\bar{\beta} \cX{\ms_0}$}\\ \cline{2-3}
	& $\dotcX{mRNA_j} = \frac{\alpha_j P^\mrm{tot}_j }{1+ \frac{K}{\cX{\ms_0}^q}}  - \gamma_j \cX{mRNA_j}$
    & $\dotcX{mRNA_j} = \frac{\alpha_j P^\mrm{tot}_j}{1+\frac{\cX{\ms_0}^q}{K}}  - \gamma_j \cX{mRNA_j}$\\\hline
  Outputs (under Assumption~\ref{as:tr-homogeneous})&\multicolumn{2}{|c|}{$y_j = \beta_j \cX{mRNA_j}$}\\ \hline
IOSCFs
(under Assumption~\ref{as:tr-homogeneous}) & 
$\displaystyle y_j^*= \frac{\alpha_j\beta_j P^\mrm{tot}}{\gamma_j\Big(1 + \frac{K}{(\theta u^*)^q}\Big)}$
 &
$\displaystyle y_j^*= \frac{\alpha_j\beta_jP^\mrm{tot}}{\gamma_j\Big(1 + \frac{(\theta u^*)^q}{K}\Big)} $ \\ \hline
LTFs
(under Assumptions~\ref{as:tr-homogeneous}--\ref{as:tr-timescale}) &
$\displaystyle H_j(s)= \frac{q K  P^\mrm{tot}\alpha_j \beta_j (\theta u^*)^{q-1}}{(K +(\theta u^*)^q)^2 (s+\gamma_j)(s+\bar{\beta})}$ &
$\displaystyle H_j(s)= -\frac{q K  P^\mrm{tot}\alpha_j\beta_j (\theta u^*)^{q-1}}{(K +(\theta u^*)^q)^2 (s+\gamma_j)(s+\bar{\beta})}$\\ \hline
        & \multicolumn{2}{|c|}{
          where $K\eqdef \frac{\koff}{\kon}, \qquad \theta =\eqdef \frac{1}{\bbeta}$}\\ \hline
\end{tabular}
}
\caption{Dynamics of a transcriptional regulation (TR) module and the
  corresponding IOSCF and LTF for each type of output. This module is
  positive and the equilibrium defined by any constant
  input $u^*$ is locally asymptotically stable, under Assumptions~\ref{as:tr-homogeneous}--\ref{as:tr-timescale}. The module is also cooperative under these assumptions when all outputs are activating.}
\label{tab:mod}
\end{table}
Figure \ref{fig:tr} shows a biological representation and the corresponding DBSR graph of a TR module. The reader may verify from Table~\ref{tab:mod} that the module satisfies Rules~\ref{ru:species}--\ref{ru:dbg} in that (i) its state contains
the concentrations of all the chemical species associated with the
module, (ii) the parameters of all the chemical reactions associated
with the module are not needed outside this module, and (iii) the
inputs and outputs of the module are rates of protein
production/degradation.

\subsection{Enzyme-substrate reaction (ESR) module}
\label{subsec:esrm}

\begin{figure}[h]
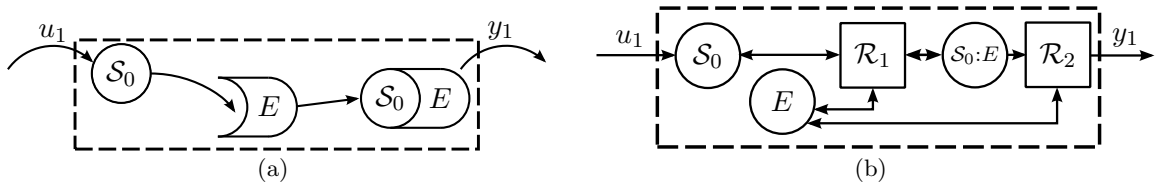

	\centering 
	{\scriptsize
		\psfrag{S0}[][]{\normalsize$\ms_0$}
		\psfrag{R1}[][]{\normalsize$\mr_1$}
		\psfrag{R2}[][]{\normalsize$\mr_2$}
		\psfrag{C}[][]{$\bndxy{\ms_0}{E}$}
		\psfrag{E}[][]{\normalsize$E$}
		\psfrag{u1}[][]{\normalsize$u_1$}
		\psfrag{y1}[][]{\normalsize$y_1$}
		\hfill\subfigure[\label{fig:bioesr}]{\includegraphics[width = 0.45\textwidth]{bioesr.ps}}\hfill
		\subfigure[\label{fig:dbsresr}]{\includegraphics[width = 0.45\textwidth]{dbsresr.ps}}\hspace*{\fill}
		\caption{\subref{fig:bioesr} Biological representation and \subref{fig:dbsresr} DBSR graph of an ESR module.}
		\label{fig:esr}
	}
\end{figure}

The \emph{enzyme-substrate reaction (ESR) module} represents the
process by which a substrate protein $\ms_0$ is covalently modified by
an enzyme $E$ into an alternative form $\ms_1$. The chemical species
associated with this module are the substrate protein $\ms_0$, the
enzyme $E$, and the complex $\bndxy{\ms_0}{E}$ formed by the
enzyme-substrate binding. The input $u$ to the ESR module is the rate
of production of the substrate $\ms_0$ due to an exogenous process
(e.g., a TR or another ESR module) and can be associated with the
generic reaction
\begin{equation*}
  \zrxn{u}{\ms_0}.
\end{equation*}
The additional reactions associated with the ESR module include the
$\ms_0$ degradation reaction
\begin{equation*}
  \orxn{\ms_0}{\gamma\cX{\ms_0}}{\varnothing},
\end{equation*}
and the reactions involved in the Michaeles-Menten model for the
enzyme-substrate interaction:
\begin{equation*}
\orxn{\ms_0 + E \xrightleftharpoons[k^\mrm{r}\cX{\bndxy{\ms_0}{E}}]{k^\mrm{f}\cX{\ms_0}\cX{E}}\bndxy{\ms_0}{E}}{k^\mrm{cat}\cX{\bndxy{\ms_0}{E}}}{\ms_1 + E},
\end{equation*}
where the total concentration of enzyme
$E^\mrm{tot}=[E]+[\bndxy{\ms_0}{E}]$ is assumed to remain constant.
The output $y$ of the ESR module is the rate of production of the
modified substrate $\ms_1$, given by
\begin{align*}
  y = k^\mrm{cat}\cX{\bndxy{\ms_0}{E}}.
\end{align*}
Table~\ref{tab:esrm} shows the system of ODEs that correspond to the
ESR module, as well as its IOSCF, and LTF. For simplicity,
instead of presenting the exact dynamics of the module (which are
straighforward to derive using MAK), we present two common
approximations to the Michaeles-Menten model:
\begin{assumption}[Equilibrium Approximation~\cite{sneyd2008mathematical}]
  \label{as:esr-equilibrium}
  The reversible reaction is in thermodynamic equilibrium (i.e., $\kf
  \cX{\ms_0}\cX{E} = \kr\cX{\bndxy{\ms_0}{E}}$), which is valid when $\kr
  \gg k^\mrm{cat}$.\frqed
\end{assumption}
\begin{assumption}[Quasi Steady-State Approximation~\cite{briggs1925note}]
  \label{as:esr-quasi}
  The concentration of the complex $\bndxy{\ms_0}{E}$ does not change
  on the timescale of product formation (i.e., $\kf \cX{\ms_0}\cX{E} =
  \kr\cX{\bndxy{\ms_0}{E}} + \kcat \cX{\bndxy{\ms_0}{E}}$), which is
  valid either when $k^\mrm{r}+k^\mrm{cat}\gg k^\mrm{f}$ or when
  $\cX{\ms_0} \gg E^{\tu{tot}}$.\frqed
\end{assumption}
Figure \ref{fig:esr} shows a biological representation and the corresponding DBSR graph of an ESR module.
\begin{table}[h]
\centering
\scalebox{0.9}{\begin{tabular}{|>{\centering\arraybackslash}m{10ex}|>{\centering\arraybackslash}m{.4\textwidth}|>{\centering\arraybackslash}m{.42\textwidth}|}
  \hline
  & Equilibrium Approximation (Assumption~\ref{as:esr-equilibrium})&
  Quasi Steady-State Approximation (Assumption~\ref{as:esr-quasi})\\ \hline
  Dynamics&
  $\dotcX{\ms_0} = \frac{u - \gamma \cX{\ms_0} - \frac{\kcat E^{\tu{tot}}\cX{\ms_0}}{K^\mrm{d} + \cX{\ms_0}}}{1 + \frac{K^\mrm{d} E^{\tu{tot}}}{(K^\mrm{d} +\cX{\ms_0}^2)}}$
  & 
  $\displaystyle \dotcX{\ms_0}
  = u - \gamma \cX{\ms_0} - \frac{k^\mrm{cat}E^{\tu{tot}}\cX{\ms_0}}{K^\mrm{m} + \cX{\ms_0}}$
  \\ \hline
  Output equation&
  $\displaystyle y = \frac{k^\mrm{cat}E^{\tu{tot}}\cX{\ms_0}}{K^\mrm{d} + \cX{\ms_0}}$, 
  $\displaystyle K^\mrm{d} \eqdef  \frac{k^\mrm{r}}{k^\mrm{f}}$ &
  $\displaystyle y = \frac{k^\mrm{cat}E^{\tu{tot}}\cX{\ms_0}}{K^\mrm{m} + \cX{\ms_0}}$,
  $\displaystyle K^\mrm{m} \eqdef \frac{k^\mrm{r}+k^\mrm{cat}}{k^\mrm{f}}$
  \\ \hline
  IOSCF & 
   $\displaystyle y^* = \frac{1}{2}(2u^*+K^\mrm{e} -\sqrt{(K^\mrm{e})^2 + 4K^\mrm{d} u^* \gamma})$,
   $K^\mrm{e} \eqdef -u^* + k^\mrm{cat}E^{\tu{tot}} + K^\mrm{d} \gamma$  & 
  $\displaystyle y^* = \frac{1}{2}(2u^*+K^\mrm{f} -\sqrt{(K^\mrm{f})^2 + 4K^\mrm{m} u^* \gamma})$,
  $K^\mrm{f} \eqdef -u^* + k^\mrm{cat}E^{\tu{tot}} + K^\mrm{m} \gamma$
  \\ \hline
  LTF &
  $\displaystyle H(s) = \frac{K^\mrm{p}(u^*)}{s + K^\mrm{q}(u^*)}$&
  $\displaystyle H(s) = \frac{K^\mrm{h}}{s+K^\mrm{h} + \gamma}$,
  $K^\mrm{h} \eqdef \frac{k^\mrm{cat}E^{\tu{tot}}K^\mrm{m}}{(K^\mrm{m} + \frac{1}{2\gamma}(K^\mrm{f} + \sqrt{(K^\mrm{f})^2 + 4K^\mrm{m} u^* \gamma}))^2}$
  \\ \hline
\end{tabular}}
\caption{Approximate dynamics of the ESR module based on the
  Equilibrium approximation \cite{sneyd2008mathematical} and the Quasi
  steady-state approximation \cite{briggs1925note}. The constants
  $K^\mrm{p}(u^*)$ and $K^\mrm{q}(u^*)$ that appear in the module LTF under Assumption~\ref{as:esr-equilibrium} are
  included only in Appendix \ref{app:modchar}, due to lack of space.  This module is
  positive, cooperative, and the equilibrium defined by any constant
  input $u^*$ is globally asymptotically stable under both assumptions.}
  \label{tab:esrm}
\end{table}

\subsection{PD-cycle module}
\label{subsec:pdcycle}
\begin{figure}[h]
  \centering
  \includegraphics[scale=.2]{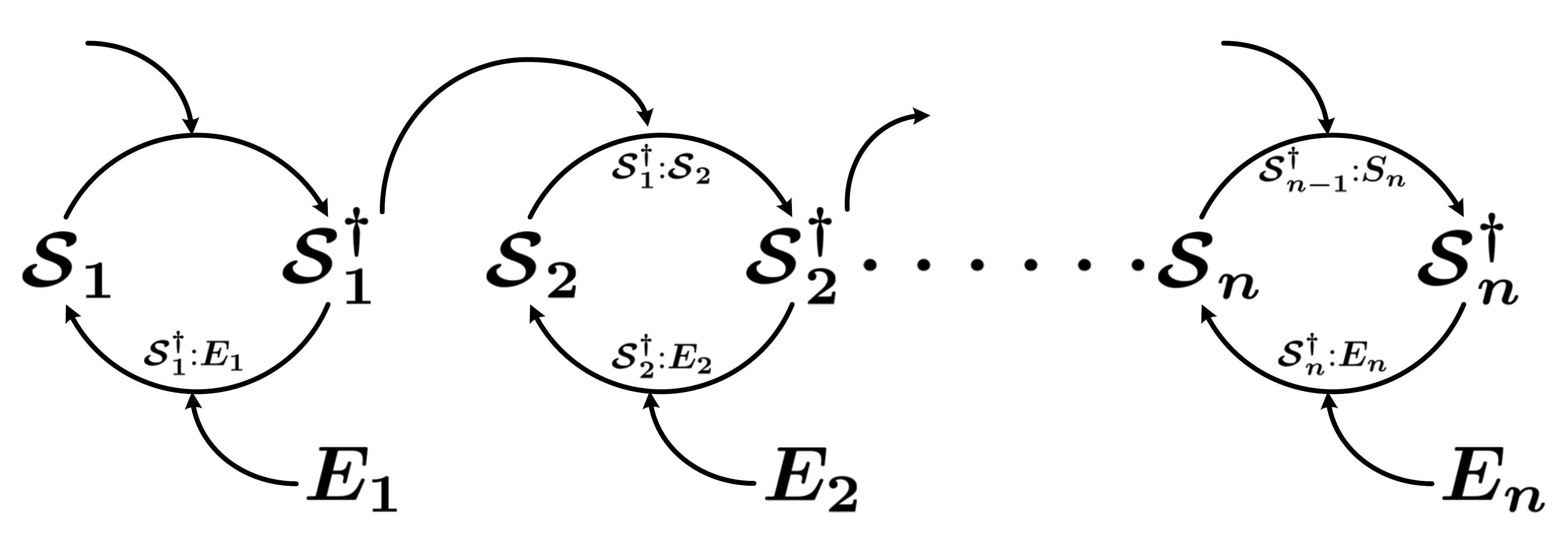}
  \caption{Representation of an $n$-stage signaling cascade, adapted
    from \cite{ossareh2011long}.}
  \label{fig:sigcasc}
\end{figure}

Signaling pathways are common networks used by cells to transmit and
receive information. A well-known signal transduction pathway, known
as a signaling cascade, consists of the series of phosphorylation and
dephosphorylation (PD) cycles shown in Figure~\ref{fig:sigcasc}.

Each of these cycles, typically referred to as a \emph{stage} in the
cascade, consists of a signaling protein that can exist in either an
inactive ($\ms_i$) or an active form ($\ms_i^\dagger$). The protein is
activated by the addition of a phosphoryl group and is inactivated by
its removal \cite{ventura2008hidden}. The activated protein
$\ms_i^\dagger$ then goes on to act as a kinase for the
phosphorylation or activation of the protein $\ms_{i+1}$ at the next
stage in the cascade. At each stage, there is also a phosphatase that
removes the phosphoryl group to deactivate the activated protein.

\medskip

The signaling cascade depicted in Figure~\ref{fig:sigcasc} can be
decomposed into $n$ \emph{PD-cycle modules}, each including the active
protein $\ms_{i}^\dagger$, the inactive protein $\ms_{i+1}$, the
complex $\bndxy{\ms_{i}^\dagger}{\ms_{i+1}}$ involved in the
activation of the protein in the next module, and the associated
chemical reactions:
\begin{align*}
  \ms_{i}^\dagger + \ms_{i+1}
  \xrightleftharpoons[k^\mrm{r}_i{[\bndxy{\ms_{i}^\dagger}{\ms_{i+1}}]}]{k^\mrm{f}_i[\ms_{i}^\dagger]\cX{\ms_{i+1}}}\bndxy{\ms_{i}^\dagger}{\ms_{i+1}}
  \xrightarrow{\alpha_i{[\bndxy{\ms_{i}^\dagger}{\ms_{i+1}}]}}  \ms_{i}^\dagger + \ms_{i+1}^\dagger.
\end{align*}
In addition, the module also includes the phosphatase enzyme
$E_{i}$, the complex $\bndxy{\ms_{i}^\dagger}{E_{i}}$ involved in the
deactivation of $\ms_{i}^\dagger$, and the
associated chemical reactions:
\begin{align*}
  \ms_{i}^\dagger + E_{i} 
  \xrightleftharpoons[\bar{k}^\mrm{r}_{i}{[\bndxy{\ms_{i}^\dagger}{E_{i}}]}]{\bar{k}^\mrm{f}_{i}[\ms_{i}^\dagger]\cX{E_{i}}}\bndxy{\ms_{i}^\dagger}{E_{i}}
  \xrightarrow{\bar{\alpha}_{i}{[\bndxy{\ms_{i}^\dagger}{E_{i}}]}}\ms_{i} + E_{i},
\end{align*}
where the total concentration of the enzyme
$E^\mrm{tot}_i=[\bndxy{\ms_{i}^\dagger}{E_{i}}]+\cX{E_{i}}$ is assumed to
remain constant.
The two inputs to this PD-cycle module are the rate of production
$u_i$ of the active protein $\ms_{i}^\dagger$ due to the activation of
$\ms_i$ in the preceding module of the cascade, and the rate of
production $v_i$ of the inactive protein $\ms_{i+1}$ due to
dephosphorylation of $\ms_{i+1}^\dagger$ in the subsequent module of the
cascade. Consistently, the two outputs of this PD-cycle are the rate
$y_i$ of production of $\ms_{i+1}^\dagger$ due to the activation of
$\ms_{i+1}$, to be used as an input to a subsequent module; and the
rate $z_i$ of production of $\ms_{i}$ due to the deactivation of
$\ms_{i}^\dagger$ to be used as an input to a preceding module.

\medskip

\begin{figure}[h]
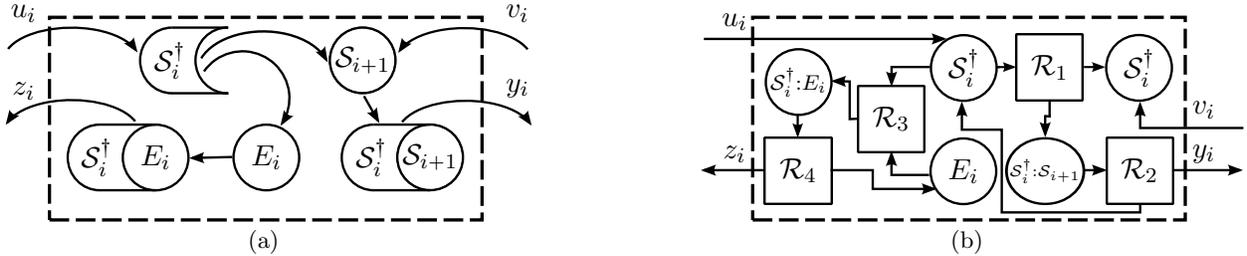

	\centering 
	{\scriptsize
		\psfrag{Sd}[][]{\normalsize$\ms_i^\dagger$}
		\psfrag{Si}[][]{\normalsize$\ms_{i+1}$}
		\psfrag{R1}[][]{\normalsize$\mr_1$}
		\psfrag{R2}[][]{\normalsize$\mr_2$}
		\psfrag{R3}[][]{\normalsize$\mr_3$}
		\psfrag{R4}[][]{\normalsize$\mr_4$}
		\psfrag{C1}[][]{$\bndxy{\ms_i^\dagger}{E_i}$}
		\psfrag{C2}[][]{\tiny$\bndxy{\ms_i^\dagger}{\ms_{i+1}}$}
		\psfrag{E}[][]{\normalsize$E_i$}
		\psfrag{ui}[][]{\normalsize$u_i$}
		\psfrag{yi}[][]{\normalsize$y_i$}
		\psfrag{vi}[][]{\normalsize$v_i$}
		\psfrag{zi}[][]{\normalsize$z_i$}
		\subfigure[\label{fig:biopdcycle}]{\includegraphics[width = 0.45\textwidth]{biopdcycle.ps}}\hfill
		\subfigure[\label{fig:dbsrpdcycle}]{\includegraphics[width = 0.45\textwidth]{dbsrpdcycle.ps}}\hfill
		\caption{\subref{fig:biopdcycle} Biological representation and \subref{fig:dbsrpdcycle} DBSR graph of a PD cycle module.}
		\label{fig:pdcycle}
	}
\end{figure}

Table~\ref{tab:pdcyc} shows the system of ODEs that correspond to the
PD-cycle module and the IOSCFs. The LTFs are straighforward to compute
and can be found in Appendix \ref{app:modchar}. While our decomposition
of the signaling cascade network satisfies Rules~\ref{ru:species}--\ref{ru:dbg} and hence enjoys both types of
modularity, this is not the case for the decompositions of the same
biochemical system found in
\cite{ventura2008hidden,OssarehV11,ossareh2011long}. Figure~\ref{fig:pdcycle} shows a biological representation and the corresponding DBSR graph of a PD cycle module.

\begin{table}[h]
  \centering
  \scalebox{0.9}{\begin{tabular}{|>{\centering\arraybackslash}m{10ex}|>{\centering\arraybackslash}m{.4\textwidth}|>{\centering\arraybackslash}m{.42\textwidth}|}
    \hline
    Dynamics &
    {\vspace{-1.5em}\small\begin{align*}
        \dotcX{\ms_{i+1}} &= v_i
        -k^\mrm{f}_i \cX{\ms_{i+1}} [\ms_{i}^\dagger]
        + k^\mrm{r}_i [\bndxy{\ms_{i}^\dagger}{\ms_{i+1}}]\\
        \dot{[\bndxy{\ms_{i}^\dagger}{\ms_{i+1}}]} &=
        k^\mrm{f}_i \cX{\ms_{i+1}} [\ms_{i}^\dagger]
        - (k^\mrm{r}_i+\alpha_i) [\bndxy{\ms_{i}^\dagger}{\ms_{i+1}}]\\
        \dot{[\ms_{i}^\dagger]}&= u_i
        -k^\mrm{f}_i \cX{\ms_{i+1}} [\ms_{i}^\dagger]
        + (k^\mrm{r}_i+\alpha_i) [\bndxy{\ms_{i}^\dagger}{\ms_{i+1}}]
        -\bar{k}^\mrm{f}_{i} [\ms_{i}^\dagger]\cX{E_{i}} \\&\qquad
        + \bar{k}^\mrm{r}_{i}(E^\mrm{tot}_i-\cX{E_{i}})\\
        \dotcX{E_{i}} &= 
        -\bar{k}^\mrm{f}_{i} [\ms_{i}^\dagger]\cX{E_{i}} + (\bar{k}^\mrm{r}_{i}+\bar{\alpha}_{i})(E^\mrm{tot}_i-\cX{E_{i}})
      \end{align*}\vspace{-1.5em}}
    \\ \hline
    Output equations & 
    {\vspace{-1.5em}\begin{align*}
        y_i &= \alpha_i[\bndxy{\ms_{i}^\dagger}{\ms_{i+1}}]\\
        z_i &= \bar{\alpha}_{i} (E^\mrm{tot}_i-\cX{E_{i}})
      \end{align*}\vspace{-1.5em}}
    \\ \hline
    IOSCF &
    {\vspace{-1.5em}\small\begin{align*}
        y_i^*&=v_i^*, &
        z_i^*&=u_i^*
      \end{align*}\vspace{-1.5em}}
    \\ \hline
  \end{tabular}}
  \caption{Dynamics of a PD-cycle module.}
  \label{tab:pdcyc}
\end{table}

\section{Interconnections between modules}
\label{sec:negfdbk}

In this section, we review the cascade, parallel, and feedback
interconnections, which are three basic mechanisms that can be
combined to obtain arbitrarily complex networks. Standard results in control systems theory allow us to
compute the IOSCF and LTF of a cascade, parallel or feedback
interconnection from the IOSCFs and LTFs of the constituent modules.

\subsection{Cascade interconnection}

\begin{figure}[h]
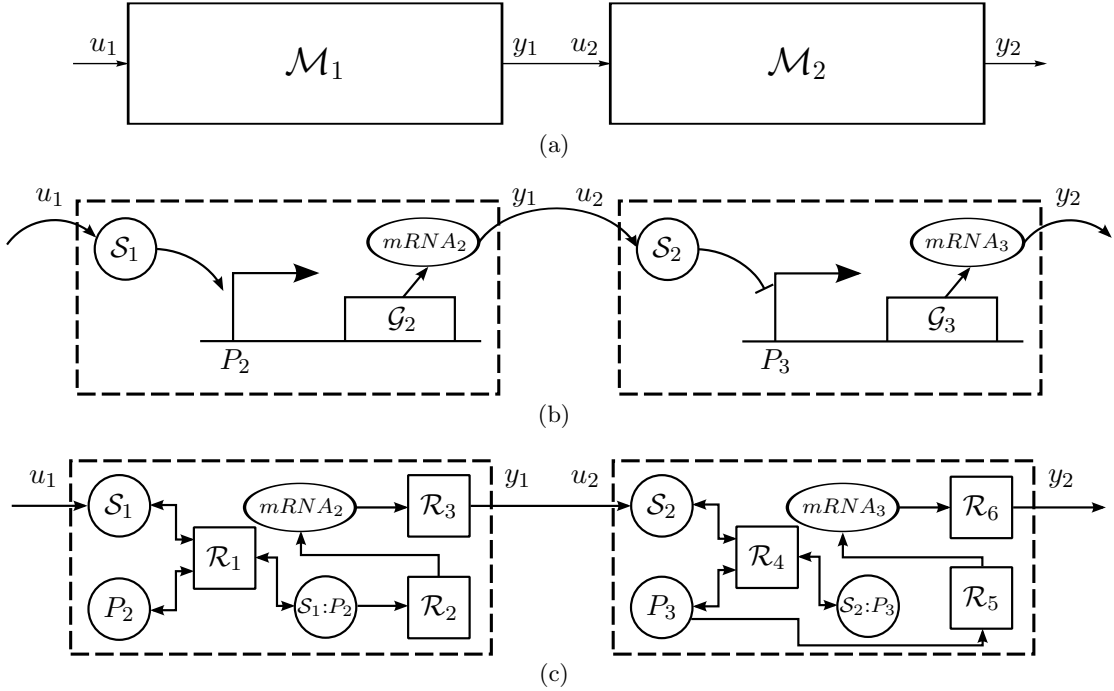

	\centering 
	{\scriptsize
		\psfrag{m2}[][]{$mRNA_2$}
		\psfrag{m3}[][]{$mRNA_3$}
		\psfrag{S1}[][]{\normalsize$\ms_1$}
		\psfrag{S2}[][]{\normalsize$\ms_2$}
		\psfrag{R1}[][]{\normalsize$\mr_1$}
		\psfrag{R2}[][]{\normalsize$\mr_2$}
		\psfrag{R3}[][]{\normalsize$\mr_3$}
		\psfrag{R4}[][]{\normalsize$\mr_4$}
		\psfrag{R5}[][]{\normalsize$\mr_5$}
		\psfrag{R6}[][]{\normalsize$\mr_6$}
		\psfrag{P2}[][]{\normalsize$P_2$}
		\psfrag{P3}[][]{\normalsize$P_3$}
		\psfrag{C1}[][]{$\bndxy{\ms_1}{P_2}$}
		\psfrag{C2}[][]{$\bndxy{\ms_2}{P_3}$}
		\psfrag{M1}[][]{\Large $\mv_1$}
		\psfrag{M2}[][]{\Large $\mv_2$}
		\psfrag{G2}[][]{\normalsize$\gene{2}$}
		\psfrag{G3}[][]{\normalsize$\gene{3}$}
		\psfrag{u1}[][]{\normalsize$u_1$}
		\psfrag{u2}[][]{\normalsize$u_2$}
		\psfrag{y1}[][]{\normalsize$y_1$}
		\psfrag{y2}[][]{\normalsize$y_2$}
		\subfigure[\label{fig:blockcasc}]{\includegraphics[width=.8\textwidth, height = 0.08\textheight]{blockcascade.ps}}\\
		\subfigure[\label{fig:biocasc}]{\includegraphics[width=.9\textwidth]{biocascade.ps}}\hfill\\
		\subfigure[\label{fig:dbsrcasc}]{\includegraphics[width=.9\textwidth]{dbsrcascade.ps}}\hfill
		  \caption{\subref{fig:blockcasc} Cascade interconnection between a
		  	module $\mv_1$ and another module $\mv_2$. \subref{fig:biocasc}
		  	Example of a cascade interconnection between a TR activator module
		  	and a TR repressor module and~\subref{fig:dbsrcasc} corresponding DBSR graph partition.}
		  \label{fig:cascexample}
	}
\end{figure}

In a \emph{cascade} interconnection between two modules, the output of
a module $\mv_1$ is connected to the input of a module $\mv_2$, as
shown in Figure~\ref{fig:blockcasc}. For example, consider the network
where a protein $\ms_1$ activates a gene $\gene{2}$, whose protein
$\ms_2$ represses a gene $\gene{3}$. This can be decomposed into a cascade
interconnection between a TR activator and a TR repressor module, as
depicted in Figures~\ref{fig:biocasc} and \ref{fig:dbsrcasc}.

When $\mv_1$ and $\mv_2$ exhibit dynamic modularity, their
input-output dynamics remain the same before and after interconnection
and the IOSCF of the cascade is given by the simple composition formula
\begin{equation}
\label{eq:cascioscf}
y_2^*=f(u_1^*)\eqdef f_2\big(f_1(u_1^*)\big),\: \forall u_1^*,
\end{equation}
where $f_1(\cdot)$ and $f_2(\cdot)$ denote the IOSCFs of the modules
$\mv_1$ and $\mv_2$, respectively.  The task of computing the
equilibrium value at the output of the cascade is therefore
straighforward once the IOSCFs of each module are
available. Similarly, the LTF of the cascade around the equilibrium
defined by the input $u_1^*$ is given by the simple multiplication formula
\begin{equation*}
H(s)=H_2(s)H_1(s),
\end{equation*}
where $H_1(s)$ and $H_2(s)$ denote the LTFs of the modules $\mv_1$ and
$\mv_2$, respectively, around the equilibrium defined by $u_1^*$ and
$f_2(u_1^*)$. The task of computing the cascade LTF is therefore
straighforward once the LTFs of each module are
available. Furthermore, $H(s)$ will be BIBO stable if both $H_1(s)$
and $H_2(s)$ are BIBO stable.

\medskip

When $\mv_1$ and $\mv_2$ admit parametric modularity, each chemical
reaction parameter is associated with a unique module. Modifying a
chemical reaction in $\mv_2$, for example, only affects the functions
$f_2(.)$ and $H_2(s)$. Consequently, the re-computation of the IOSCF
and LTF of the cascade after this modification is simplified, as
$f_1(.)$ and $H_1(s)$ remain unchanged.

\subsection{Parallel interconnection}

\begin{figure}[h]
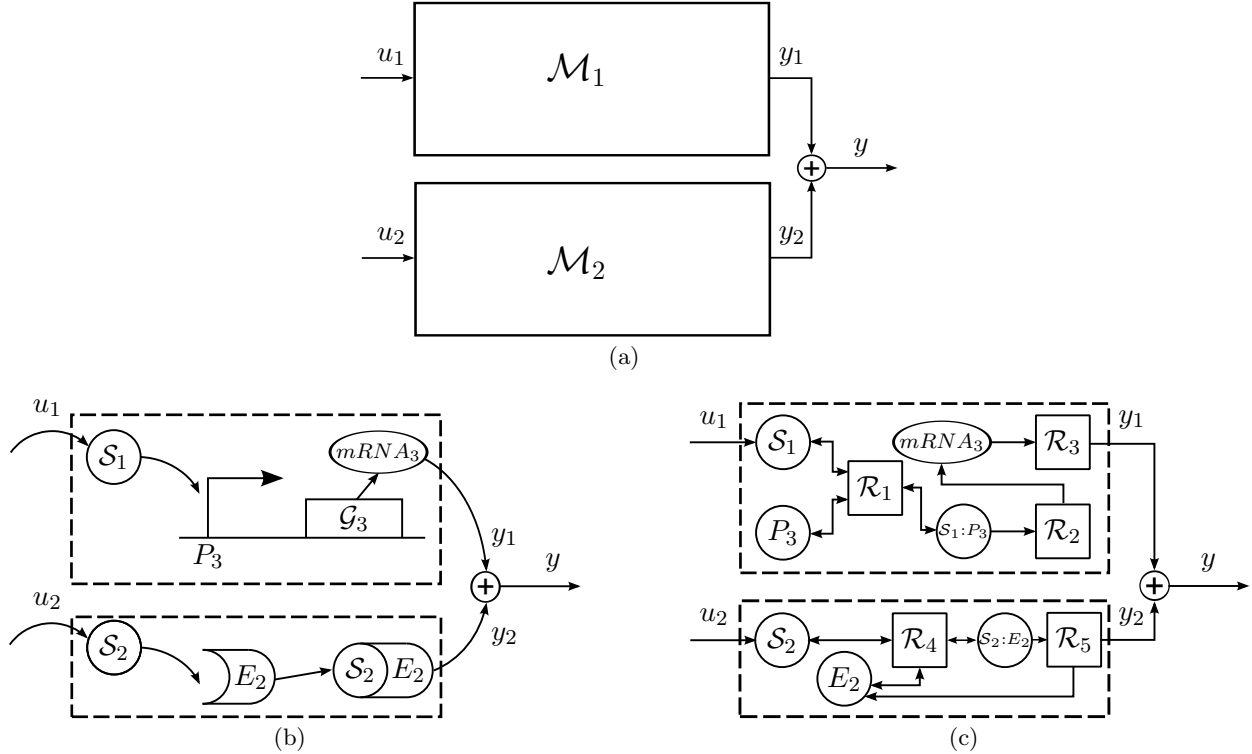

	\centering 
	{\scriptsize
		\psfrag{m3}[][]{$mRNA_3$}
		\psfrag{S1}[][]{\normalsize$\ms_1$}
		\psfrag{S2}[][]{\normalsize$\ms_2$}
		\psfrag{R1}[][]{\normalsize$\mr_1$}
		\psfrag{R2}[][]{\normalsize$\mr_2$}
		\psfrag{R3}[][]{\normalsize$\mr_3$}
		\psfrag{R4}[][]{\normalsize$\mr_4$}
		\psfrag{R5}[][]{\normalsize$\mr_5$}
		\psfrag{P2}[][]{\normalsize$P_2$}
		\psfrag{P3}[][]{\normalsize$P_3$}
		\psfrag{C1}[][]{\tiny$\bndxy{\ms_1}{P_3}$}
		\psfrag{C2}[][]{\tiny$\bndxy{\ms_2}{E_2}$}
		\psfrag{M1}[][]{\Large $\mv_1$}
		\psfrag{M2}[][]{\Large $\mv_2$}
		\psfrag{E2}[][]{\normalsize$E_2$}
		\psfrag{G3}[][]{\normalsize$\gene{3}$}
		\psfrag{u1}[][]{\normalsize$u_1$}
		\psfrag{u2}[][]{\normalsize$u_2$}
		\psfrag{y1}[][]{\normalsize$y_1$}
		\psfrag{y2}[][]{\normalsize$y_2$}
		\psfrag{y}[][]{\normalsize$y$}
		\subfigure[\label{fig:blockpar}]{\includegraphics[width=.45\textwidth, height = 0.2\textheight]{blockparallel.ps}}\hfill\\
		\subfigure[\label{fig:biopar}]{\includegraphics[width=.47\textwidth]{bioparallel.ps}}\hfill
		\subfigure[\label{fig:dbsrpar}]{\includegraphics[width=.47\textwidth]{dbsrparallel.ps}}\hfill
	\caption{\subref{fig:blockpar} Parallel interconnection of two
	modules $\mv_1$ and $\mv_2$. \subref{fig:biopar} Example of a
	parallel interconnection between an activator TR module and an ESR module and~\subref{fig:dbsrpar} corresponding DBSR graph partition.}
	}
	\label{fig:parexamples}
\end{figure}

In a \emph{parallel} interconnection between the modules $\mv_1$ and
$\mv_2$, the outputs of the modules are summed as shown in
Figure~\ref{fig:blockpar}. For example, consider a network where a
protein $\ms_1$ activates a gene $\gene{3}$ that produces $\ms_3$, and
simultaneously the same protein $\ms_3$ can be produced by a covalent
modification of another protein $\ms_2$ by an enzyme $E_2$. This network can be
decomposed into the parallel interconnection of a TR activator module
and an ESR module, as depicted in Figures~\ref{fig:biopar} and \ref{fig:dbsrpar}, where $\ms_1$ and $\ms_2$ are produced by some
exogenous process at rates $u_1$ and $u_2$, respectively.

Under dynamic modularity, the (two-input) IOSCF of the parallel
interconnection is given by the simple addition formula
\begin{equation*}
  u_3^*=f(u_1^*,u_2^*)\eqdef f_1(u_1^*)+f_2(u_2^*), \:\forall u_1^*,u_2^*,
\end{equation*}
where $f_1(\cdot)$ and $f_2(\cdot)$ denote the IOSCFs of the modules
$\mv_1$ and $\mv_2$, respectively.  Furthermore, the (two-input) LTF
of the parallel interconnection around the equilibrium defined by the
input pair $(u_1^*,u_2^*)$ is obtained by stacking $H_1(s)$, $H_2(s)$
side-by-side, as in
\begin{equation*}
H(s)=[H_1(s) \; H_2(s)],
\end{equation*}
where $H_1(s)$ and $H_2(s)$ denote the LTFs of the modules $\mv_1$ and
$\mv_2$, respectively, around the equilibria defined by $u_1^*$ and
$u_2^*$. Also here, $H(s)$ will be BIBO stable if both $H_1(s)$ and
$H_2(s)$ are BIBO stable. Furthermore, parametric modularity ensures
that modifying a chemical reaction associated with $\mv_i$,
$i\in\{1,2\}$ only affects the functions $f_i(.)$ and $H_i(s)$.

\subsection{Feedback interconnection}
\label{subsec:fbk}
In a \emph{feedback} interconnection, the output of a module $\mv_1$
is connected back to its input through a summation block, as shown in
Figure~\ref{fig:blockfeed}. An example of a feedback interconnection is an
auto-repressor, where a protein $\ms_1$, which is produced by an
exogenous process at a rate $u$, represses its own gene $\gene{1}$,
as illustrated in Figure~\ref{fig:biofeed}.

\begin{figure}[h]
	\centering 
	{\scriptsize
		\psfrag{m1}[][]{\tiny$mRNA_1$}
		\psfrag{S1}[][]{\normalsize$\ms_1$}
		\psfrag{S2}[][]{\normalsize$\ms_2$}
		\psfrag{R1}[][]{\normalsize$\mr_1$}
		\psfrag{R2}[][]{\normalsize$\mr_2$}
		\psfrag{R3}[][]{\normalsize$\mr_3$}
		\psfrag{P1}[][]{\normalsize$P_1$}
		\psfrag{C1}[][]{\tiny$\bndxy{\ms_1}{P_1}$}
		\psfrag{M1}[][]{\Large $\mv_1$}
		\psfrag{G1}[][]{\normalsize$\gene{1}$}
		\psfrag{u}[][]{\normalsize$u$}
		\psfrag{u1}[][]{\normalsize$u_1$}
		\psfrag{u2}[][]{\normalsize$u_2$}
		\psfrag{y1}[][]{\normalsize$y_1$}
		\psfrag{y2}[][]{\normalsize$y_2$}
		\psfrag{y}[][]{\normalsize$y$}
		\subfigure[\label{fig:blockfeed}]{\includegraphics[width=.45\textwidth, height = 0.1\textheight]{blockfeedback.ps}}\hfill\\
		\subfigure[\label{fig:biofeed}]{\includegraphics[width=.47\textwidth]{biofeedback.ps}}\hfill
		\subfigure[\label{fig:dbsrfeed}]{\includegraphics[width=.47\textwidth]{dbsrfeedback.ps}}\hfill
		\caption{\subref{fig:blockfeed} Feedback interconnection of module $\mv_1$. \subref{fig:biopar} Example of a feedback interconnection of a TR repressor module and ~\subref{fig:dbsrpar} corresponding DBSR graph partition.}
	}
	\label{fig:feedexamples}
\end{figure}

When $\mv_1$ exhibits dynamic modularity, the IOSCF of the feedback
interconnection is given by $f(u^*)=y^*$, where $y^*$ is the solution
to the equation
\begin{align*}
y^*&=f_1(u^*+y^*),
\end{align*}
where $f_1(\cdot)$ denotes the IOSCF of the module $\mv_1$.  When this
equation has multiple solutions, the IOSCF of the feedback is not well
defined. When the IOSCF is well-defined, the LTF of the feedback
interconnection around the equilibrium defined by the input $u^*$ is
given by
\begin{equation*}
H(s)=\big(I-H_1(s)\big)^{-1}H_1(s)=H_1(s)\big(I-H_1(s)\big)^{-1}.
\end{equation*}
where $H_1(s)$ denote the LTFs of the module $\mv_1$ around the
equilibria defined by $u^*+f(u^*)$, and $I$ represents the identity
matrix of the same size as the matrix $H_1(s)$.

\medskip

The interconnection in Figure~\ref{fig:blockfeed} is said to
correspond to a \emph{positive (negative) feedback} if the module
$\mv_1$ has a well-defined monotone increasing (decreasing) IOSCF with
respect to each of its inputs, which means that an increase in the
constant input $u_1^*$ results in an increase (decrease) in the
equilibrium output $y_1^*$.

\medskip

For the cascade and parallel interconnections, the IOSCF of the
networks were well-defined provided that the IOSCFs of the constituent
modules were well-defined. Moreover, if the constituent modules had
BIBO stable LTFs, the LTF of the interconnection was also BIBO
stable. This is no longer true for a feedback interconnection; a
module $\mv_1$ may have a well-defined IOSCF and a BIBO stable LTF,
but the feedback interconnection of $\mv_1$ may have multiple
equilibria, and the LTFs around these equilibria may or may not be
BIBO stable, thus requiring more sophisticated tools to study feedback
interconnections.

\subsection{Nested interconnection structures}
\label{subsec:nest}

\begin{figure}[h]
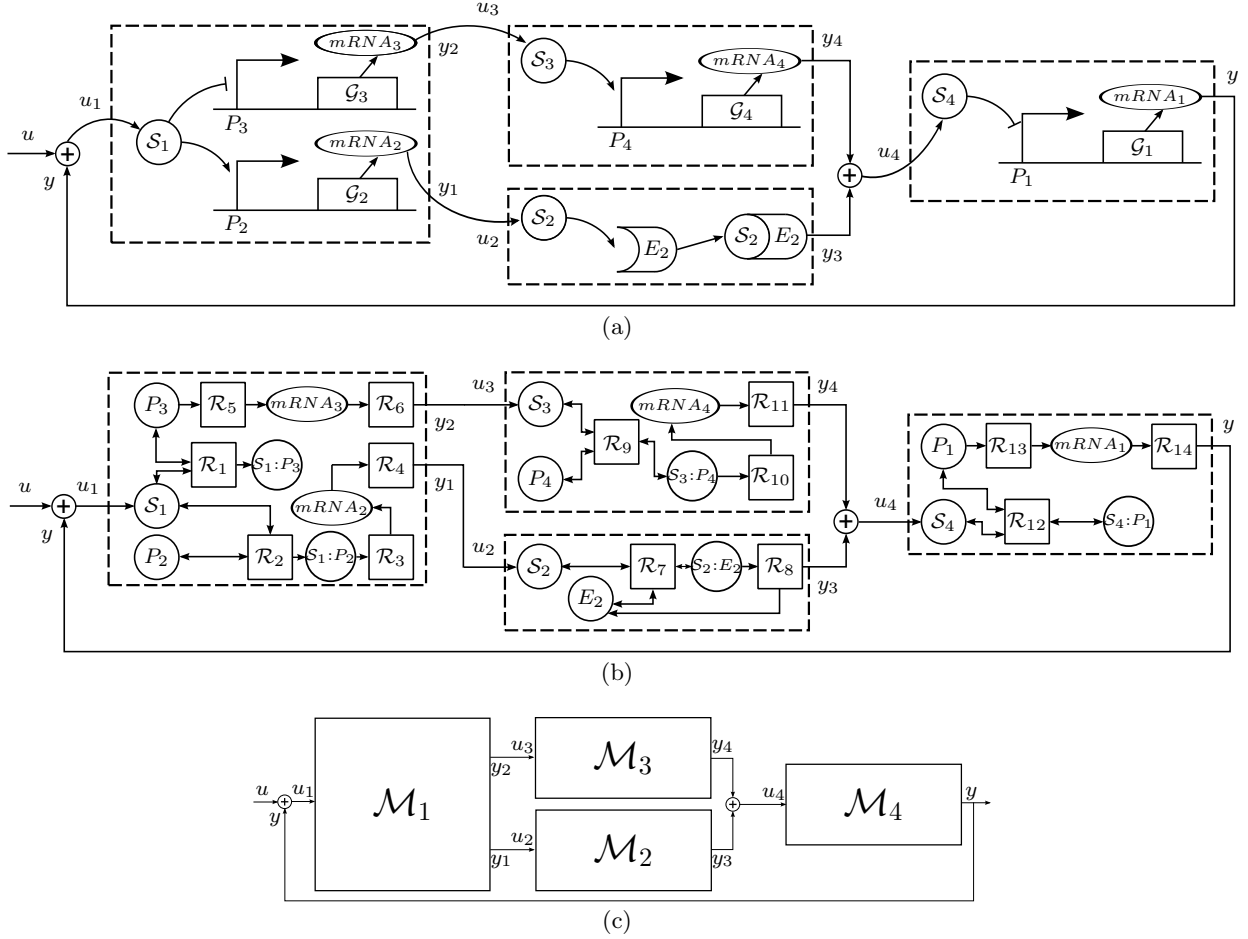

\centering
{\scriptsize
\psfrag{S1}[][]{$\ms_1$}
\psfrag{S2}[][]{$\ms_2$}
\psfrag{S3}[][]{$\ms_3$}
\psfrag{S4}[][]{$\ms_4$}
\psfrag{m1}[][]{\tiny$mRNA_1$}
\psfrag{m2}[][]{\tiny$mRNA_2$}
\psfrag{m3}[][]{\tiny$mRNA_3$}
\psfrag{m4}[][]{\tiny$mRNA_4$}
\psfrag{E2}[][]{$E_2$}
\psfrag{M1}[][]{\Large $\mv_1$}
\psfrag{M2}[][]{\Large $\mv_2$}
\psfrag{M3}[][]{\Large $\mv_3$}
\psfrag{M4}[][]{\Large $\mv_4$}
\psfrag{G1}[][]{$\gene{1}$}
\psfrag{G2}[][]{$\gene{2}$}
\psfrag{G3}[][]{$\gene{3}$}
\psfrag{G4}[][]{$\gene{4} $}
\psfrag{P1}[][]{$P_{1}$}
\psfrag{P2}[][]{$P_{2}$}
\psfrag{P3}[][]{$P_{3}$}
\psfrag{P4}[][]{$P_{4}$}
\psfrag{C1}[][]{\tiny$\bndxy{\ms_1}{P_2}$}
\psfrag{C2}[][]{\tiny$\bndxy{\ms_1}{P_3}$}
\psfrag{C3}[][]{\tiny$\bndxy{\ms_2}{E_2}$}
\psfrag{C4}[][]{\tiny$\bndxy{\ms_3}{P_4}$}
\psfrag{C5}[][]{\tiny$\bndxy{\ms_4}{P_1}$}
\psfrag{R1}[][]{$\mr_{1}$}
\psfrag{R2}[][]{$\mr_{2}$}
\psfrag{R3}[][]{$\mr_{3}$}
\psfrag{R4}[][]{$\mr_{4}$}
\psfrag{R5}[][]{$\mr_{5}$}
\psfrag{R6}[][]{$\mr_{6}$}
\psfrag{R7}[][]{$\mr_{7}$}
\psfrag{R8}[][]{$\mr_{8}$}
\psfrag{R9}[][]{$\mr_{9}$}
\psfrag{R10}[][]{$\mr_{10}$}
\psfrag{R11}[][]{$\mr_{11}$}
\psfrag{R12}[][]{$\mr_{12}$}
\psfrag{R13}[][]{$\mr_{13}$}
\psfrag{R14}[][]{$\mr_{14}$}
\psfrag{u1}[][]{$u_1$}
\psfrag{u2}[][]{$u_2$}
\psfrag{u3}[][]{$u_3$}
\psfrag{u4}[][]{$u_4$}
\psfrag{u}[][]{$u$}
\psfrag{y1}[][]{$y_1$}
\psfrag{y2}[][]{$y_2$}	
\psfrag{y3}[][]{$y_3$}	
\psfrag{y4}[][]{$y_4$}	
\psfrag{y}[][]{$y$}	
\subfigure[\label{fig:bionested}]{\includegraphics[width=\textwidth]{bionested.ps}}\\
\subfigure[\label{fig:dbsrnested}]{\includegraphics[width=\textwidth]{dbsrnested.ps}}
\subfigure[\label{fig:blocknested}]{\includegraphics[width=.6\textwidth]{blocknested.ps}}\\
\caption{\subref{fig:bionested} Biological representation of the nested interconnection structure and \subref{fig:dbsrnested} corresponding DBSR graph partition. \subref{fig:blocknested} Block diagram representation of network decomposition, which shows $\mv_1$ in cascade with the parallel interconnection of $\mv_2$ and $\mv_3$, connected in cascade with $\mv_4$ and in turn, connected in a feedback loop.}
\label{fig:nestexamples}
}
\end{figure}

In general, arbitrarily complex biological networks can be decomposed
into combinations of cascade, parallel and feedback interconnection
structures, also known as \emph{nested} interconnection
structures. One such example, involving TR and ESR modules, is
depicted in Figure~\ref{fig:nestexamples}. In this case, a protein $\ms_1$, which is produced by some exogenous process at a rate $u$, activates a gene $\gene{2}$ while simultaneously repressing $\gene{3}$. The protein $\ms_2$ is then covalently modified to $\ms_4$, while $\ms_3$ activates $\gene{4}$. The protein $\ms_4$ goes on to repress $\gene{1}$, completing the feedback loop.

\subsection{Illustrative examples}

For the remainder of this section, we demonstrate how a modular
approach can be used to simplify the analysis of and provide insight
into the behavior of two well-studied biological networks that can be
represented as nested interconnection structures: the Covalent
Modification network and the Repressilator network, both of which can
be decomposed into a cascade of modules that are subsequently
connected in a feedback loop. We use this modular decomposition to
simplify the process of finding the number of equilibrium points of
these networks, and to determine whether the protein concentrations
within the modules will converge to these points.

\subsubsection*{Covalent Modification network}
\begin{figure}[h]
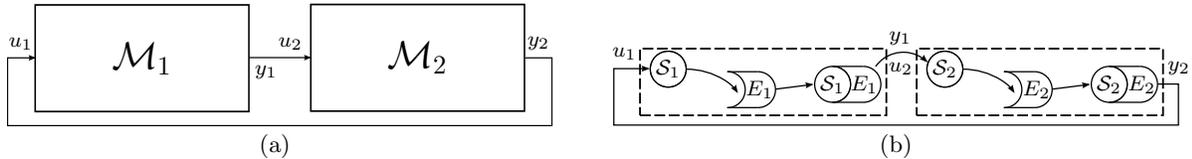

\centering 
\scriptsize{
\psfrag{S1}[][]{$\ms_1$}
\psfrag{S2}[][]{$\ms_2$}
\psfrag{E1}[][]{$E_1$}
\psfrag{E2}[][]{$E_2$}
\psfrag{M1}[][]{\Large $\mv_1$}
\psfrag{M2}[][]{\Large $\mv_2$}
\psfrag{u1}[][]{$u_1$}
\psfrag{u2}[c][l]{$u_2$}
\psfrag{y1}[c][r]{$y_1$}
\psfrag{y2}[][]{$y_2$}
\subfigure[\label{fig:blockcovmod}]{\includegraphics[width=.45\textwidth, height = 0.08\textheight]{blockcovmod.ps}}\hfill
\subfigure[\label{fig:biocovmod}]{\includegraphics[width=.47\textwidth]{biocovmod.ps}}\hspace*{\fill}
\caption{\subref{fig:blockcovmod} Block diagram representation and \subref{fig:biocasc} biological representation of a Covalent Modification network decomposition.}
\label{fig:covmodi}
}
\end{figure}

The Covalent Modification network adapted from
\cite{goldbeter1981amplified} consists of a protein that can exist in
the unmodified form $\ms_1$, or in the modified form $\ms_2$. The
interconversion between the forms is catalyzed by two enzymes $E_1$
and $E_2$. This network can be decomposed into a cascade of two ESR
modules interconnected in (positive) feedback, as shown in
Figure~\ref{fig:covmodi}. This particular
feedback interconnection does not have any exogenous input.

Based on what we saw before regarding the IOSCFs of cascade and
feedback interconnections, the value of $u_1$ at the equilibrium point of
the network must satisfy the equation 
\begin{align}  \label{eq:sigeq}
  f_2\big(f_1(u_1^*)\big) = u_1^*,
\end{align}
where $f_1(\cdot)$ and $f_2(\cdot)$ denote the IOSCFs of the two ESR
modules (see Table~\ref{tab:esrm}). Observing that $f_1(.)$ and
$f_2(.)$ --- and therefore their composition $f_2(f_1(.))$ --- are all
zero-at-zero, monotone increasing and strictly concave functions, we
can conclude that \eqref{eq:sigeq} has the unique solution $u_1^*=0$, corresponding to both substrate concentrations being $0$, \emph{for all positive values of the module parameters.}

\medskip

To determine whether the concentrations of all species converge to the
unique equilibrium point, we can apply powerful results from
\cite{angeli2004interconnections} to determine whether a network is
cooperative, based on the properties of its constituent blocks. In
particular, these results allow us to conclude that the cascade of the
two ESR blocks is cooperative (and actually enjoys a few additional
related properties, including excitability and transparency). We can
then use results from \cite{angeli2004multi} to establish that since the IOSCFs always satisfy
\begin{equation*}
\PDeriv{f_2\left(f_1\left(u_1^*\right)\right)}{u_1^*}\Big|_{u_1^*=0} <1,
\end{equation*}
the substrate concentrations converge to the unique equilibrium point $0$,
\emph{regardless of their initial concentration and of the module
  parameters}. A more detailed explanation and formal proof is provided in Appendix~\ref{app:posfbk}.

\subsubsection*{Repressilator network}

\begin{figure}[h]
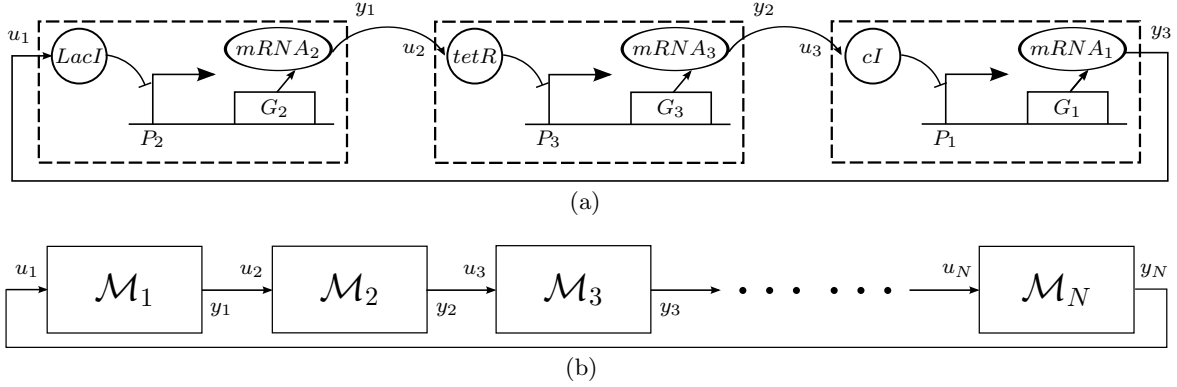

	\centering 
	{\scriptsize
		\psfrag{L}[][]{$LacI$}
		\psfrag{T}[][]{$tetR$}
		\psfrag{C}[][]{$cI$}
		\psfrag{P1}[][]{$P_1$}
		\psfrag{P2}[][]{$P_2$}
		\psfrag{P3}[][]{$P_3$}
		\psfrag{G1}[][]{$G_1$}
		\psfrag{G2}[][]{$G_2$}
		\psfrag{G3}[][]{$G_3$}
		\psfrag{M1}[][]{\Large $\mv_1$}
		\psfrag{M2}[][]{\Large $\mv_2$}
		\psfrag{M3}[][]{\Large $\mv_3$}
		\psfrag{MN}[][]{\Large $\mv_N$}
		\psfrag{m1}[][]{$mRNA_1$}
		\psfrag{m2}[][]{$mRNA_2$}
		\psfrag{m3}[][]{$mRNA_3$}
		\psfrag{u1}[][]{$u_1$}
		\psfrag{u2}[][]{$u_2$}
		\psfrag{u3}[][]{$u_3$}
		\psfrag{uN}[][]{$u_N$}
		\psfrag{y1}[][]{$y_1$}
		\psfrag{y2}[][]{$y_2$}
		\psfrag{y3}[][]{$y_3$}
		\psfrag{yN}[][]{$y_N$}
	\subfigure[\label{fig:repressilator}]{\includegraphics[width=.95\textwidth]{biorepressilator.ps}} \hfill
	\subfigure[\label{fig:fbk_spc}]{\includegraphics[width=.95\textwidth]{blockrepressilator.ps}} \hfill
	\caption{\subref{fig:repressilator} Biological realization of
          a Repressilator network consisting of three naturally-occurring repressor proteins $LacI$, $tetR$ and $cI$, each corresponding to a TR repressor module. \subref{fig:fbk_spc} Cascade of $N$ TR repressor
          modules connected in feedback, where each module is denoted
          by $\mv_i$, $i\in\{1,\cdots,N\}$.}
    }
\end{figure}

The Repressilator is a synthetic network designed to gain insight into
the behavior of biological oscillators \cite{Elowitz2000aa}. This
network consists of an odd number $N$ of repressor proteins
$\ms_1,\cdots,\ms_N$, where $\ms_i$ represses the gene $\gene{i+1}$,
for $i \in \{1,\cdots,N-1\}$ and $\ms_N$ represses the gene
$\gene{1}$. These networks can be decomposed into a cascade of $N$
single-gene TR repressor modules connected in a (negative)
feedback loop with no exogenous input. The biological realization of this decomposition with
$N=3$ (as in \cite{Elowitz2000aa}) is shown in
Figure~\ref{fig:repressilator}, with the corresponding block diagram
representation depicted in Figure~\ref{fig:fbk_spc}.

\medskip

Again based on what we saw before regarding the IOSCFs of cascade and
feedback interconnections, we conclude that the equilibrium point of
the network must satisfy the equation
\begin{align}\label{eq:f-n}
  f_N\big(\cdots f_2\big(f_1(u_1^*)\big)\cdots\big) = u_1^*,
\end{align}
where $f_i(\cdot)$ denotes the IOSCF of the $i$th TR module (see
Table~\ref{tab:mod}). Since each $f_i(\cdot)$ is monotone decreasing,
the composition of the $N$ (odd) functions is also monotone decreasing
and we have a feedback interconnection with a unique solution $u_1^*$
to \eqref{eq:f-n} \emph{for all values of the parameters.} For
simplicity, in the remainder of this section we assume that the
parameters of the chemical reactions within each TR repressor module
are exactly the same, implying that each module is identical. However,
it is straightforward to modify the results below for networks
consisting of non-identical modules.

\medskip

\begin{figure}[h]
  \centering \subfigure[Nyquist plots of the LTF of a TR repressor
  module which satisfies Assumption~\ref{as:tr-timescale} for two different
  sets of parameters: $\beta = \bar{\beta}= 10^{-1.5}$ (solid) and
  $\beta = \bbeta = 1$ (dashed). The remaining parameters are the same
  for both plots: $Ptot=1$, $K=100$, $\gamma=1$, $q= 2$, $\alpha=100$.
  The points $e^{j \frac{2\pi \ell}{3}}$, $j\in\{1,2,3\}$ that
  appear in the criterion \eqref{eq:Nyquist-N} are marked with ``X''. For the solid
  Nyquist plot, there are no clockwise encirclements of these three
  points so we have BIBO stability for the feedback LTF, whereas for
  the dashed Nyquist plot this is not the
  case.\label{fig:nyquist1}]{\includegraphics[width =
    0.44\columnwidth]{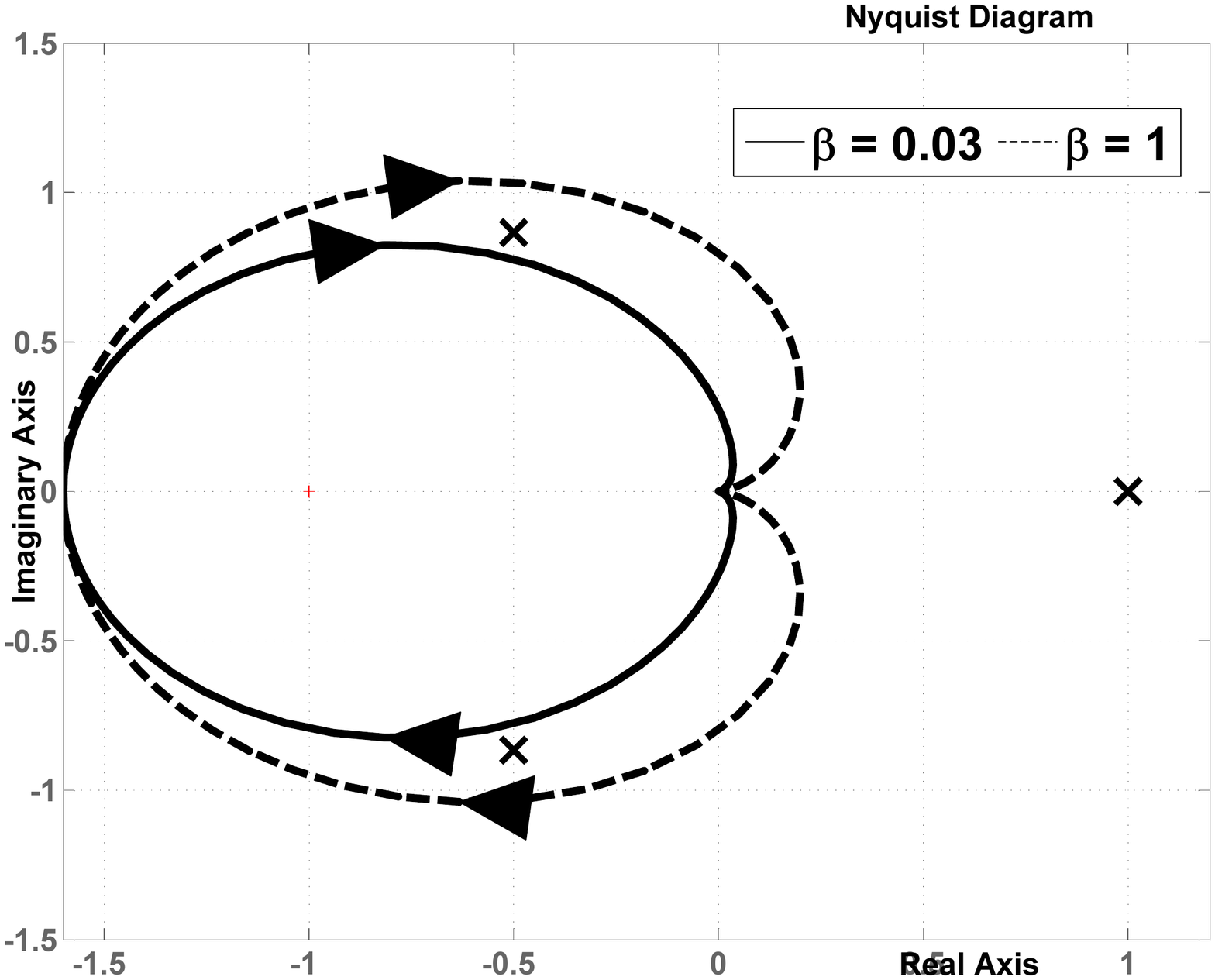}}\quad
  \subfigure[Stability regions for the 3-gene Repressilator, as a
  function of the parameters $\alpha$ and $\beta = \bar{\beta}$, under Assumption \ref{as:tr-timescale}. The
  (sufficient) condition~\eqref{eq:cond2} allows us to conclude that the equilibrium is GAS in region ``A;'' the (sufficient)
  condition~\eqref{eq:cond1} allows us to conclude that the equilibrium is LAS in
  regions ``A'' and ``B;'' and the (necessary and sufficient)
  condition \eqref{eq:Nyquist-N} allows us to conclude that the equilibrium is LAS
  in regions ``A,'' ``B'' and ``C;'' and also that it is unstable
  in region ``D.'' The Repressilator oscillates in region ``D'', assuming that there are no repeated poles.
  \cite{el2005repressilators}.\label{fig:results}]{\includegraphics[scale
    = 0.25]{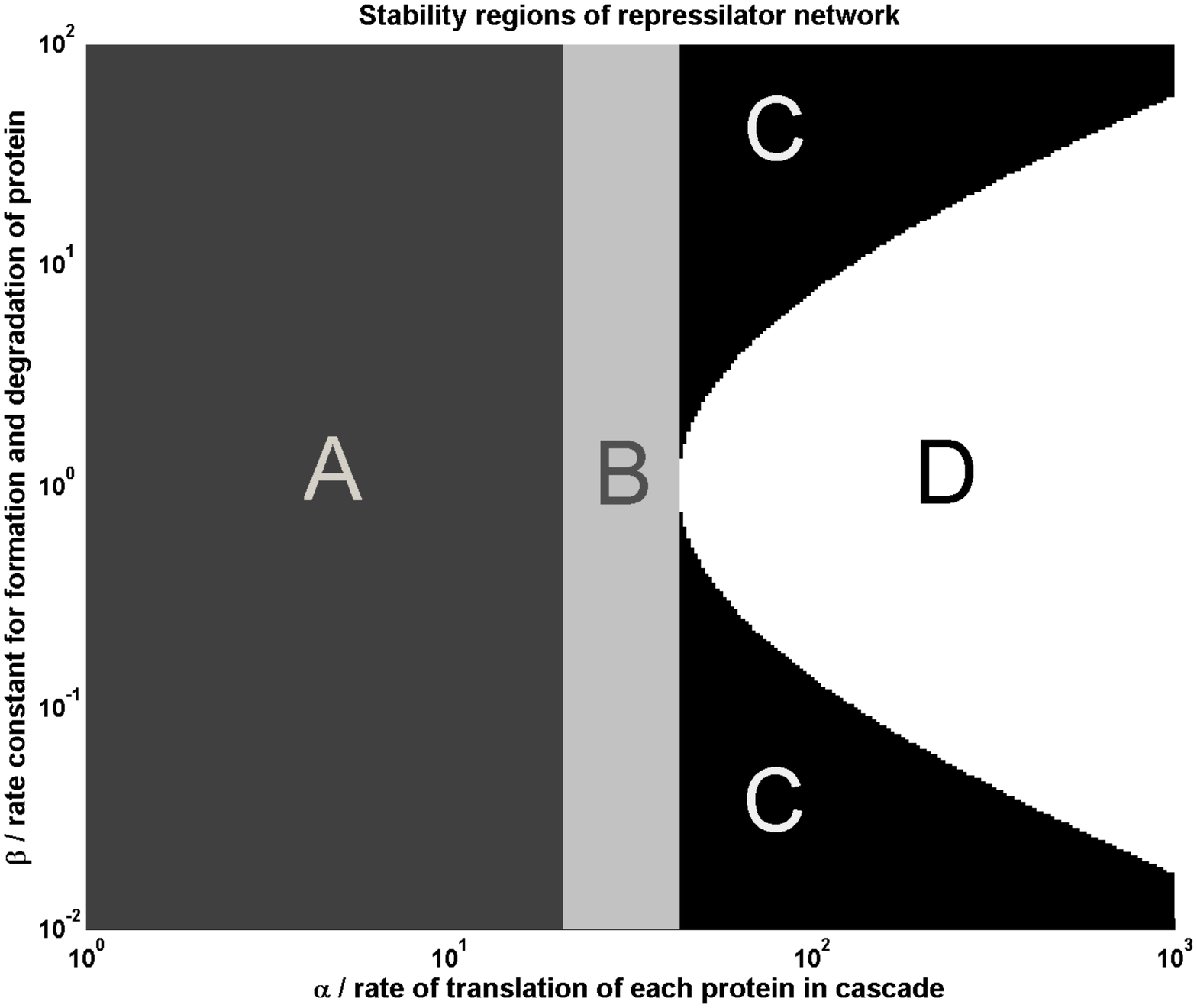}}
\caption{Nyquist plots and simulation results}
\label{fig:fullresults}
\end{figure}

For this example, we follow two alternative approaches to determine
whether or not the concentrations of the species converge to the
unique equilibrium. The first approach is based on the \emph{Nyquist
  Stability Criterion} \cite{dorfbook} and will allow us to determine
whether trajectories that start close to the equilibrium eventually
converge to it. The \emph{Nyquist plot} of a transfer function $H(s)$
is a plot of $H(j\omega)$ on the complex plane, as the (real-valued)
variable $\omega$ ranges from $-\infty$ to $+\infty$.  The plot is
labeled with arrows indicating the direction in which $\omega$
increases. Figure~\ref{fig:nyquist1} depicts the Nyquist
plot of the LTF of a single-gene TR repressor under Assumption \ref{as:tr-timescale}. According
to the Nyquist Stability Criterion, the feedback interconnection of a
cascade of $N$ modules with a BIBO stable LTF $H(s)$ has a BIBO stable
LTF if and only if
\begin{align}\label{eq:Nyquist-N}
  \sum_{\ell=1}^{N} \#END[e^{j \frac{2\pi \ell}{N}}]=0
\end{align}
where $\#END[e^{j \frac{2\pi \ell}{N}}]$ denotes the number of
clockwise encirclements of the Nyquist contour of $H(j \omega)$,
$\omega\in\R$ around the point $e^{j \frac{2\pi\ell}{N}}$ on the
complex plane. BIBO stability of the LTF generically implies that the
equilibrium point is locally asymptotically stable (LAS), which means
that trajectories that start close to the equilibrium eventually
converge to it. Figure~\ref{fig:nyquist1} shows the Nyquist plots of the LTF of a TR repressor module for two sets of parameter values: one
resulting in a Nyquist plot that satisfies \eqref{eq:Nyquist-N} and
therefore corresponds to a LAS equilibrium and the other corresponding
to an unstable equilibrium, assuming $N=3$. The Nyquist Stability Criterion therefore allows us to determine the local asymptotic stability of the equilibrium point of the overall Repressilator network from the Nyquist plot of a single TR repressor module alone.

An alternative approach that can be used to determine convergence to
the equilibrium of a negative feedback interconnection is based on the
\emph{Secant Criterion} \cite{tyson1978dynamics,thron1991secant} and
some of its more recent variations \cite{arcak2006diagonal}, which
apply to systems of the form
\begin{align}
  \label{eq:module}
  [\dot{\bar{\ms}}_i] &= u_i - c_i([\bar \ms_i]), & y_i &= d_i([\bar \ms_i]),
  & i\in\{1,2,\dots,M\},
\end{align}
with all the $d_i(\cdot)$ monotone strictly increasing, connected in
feedback according to
\begin{align*}
  u_1 &= y_M, & u_i &= y_{i-1}, \quad i \in \{2,3,\dots,M\}.
\end{align*}
When an odd number of the functions $c_i(.)$ are monotone strictly
decreasing and the remaining $c_i(\cdot)$ are monotone strictly
increasing, the unique equilibrium point of the resulting negative
feedback interconnection is LAS
if
\begin{align}
  \label{eq:cond1}
  \prod_{i=1}^{M} \bigg|\frac{\PDeriv{d_i(s_i)}{s_i}\big|_{s_i=[\bar \ms_i]^*}}{\PDeriv{c_i(s_i)}{s_i}\big|_{s_i=[\bar\ms_i]^*}}\bigg|
  < \sec\left({\frac{\pi}{M}}\right)^M
\end{align}
where the $[\bar \ms_i]^*$ denote the concentration of the species
$\bar\ms_i$ at the equilibrium. Equation \eqref{eq:cond1} is a
\emph{sufficient} but not necessary condition for local asymptotic
stability, which means that the equilibrium could still be LAS if this condition fails, as opposed to \eqref{eq:Nyquist-N} which
is a necessary and sufficient condition.

The secant-like criterion adapted from \cite{arcak2006diagonal} provides the
following sufficient (but typically not necessary) condition for the equilibrium point to be globally asymptotically stable (GAS): There exist
constants $\phi_i$, $i\in\{1,2,\dots,M\}$, such that
\begin{align}
  \label{eq:cond2}
  \prod_{i=1}^{M}\phi_i &< \sec\left({\frac{\pi}{M}}\right)^M, &
  \Big|\PDeriv{d_i(s_i)}{s_i}\Big| 
  &\leq \phi_i\PDeriv{c_i(s_i)}{s_i},\quad \forall s_i \neq \cX{\bar\ms_i}^*,\;i\in\{1,2,\dots,M\}.
\end{align}
When the equilibrium is GAS, we can conclude that the concentration of the species
converge to the unique equilibrium point, \emph{regardless of their
  initial concentration}. For either condition, knowledge of the
derivatives of the functions $c_i(.)$ and $d_i(.)$ suffices to
establish local or global asymptotic stability of the equilibrium of
the overall network.

\medskip

Under Assumption \ref{as:tr-timescale}, it turns out that each TR repressor module under corresponds to a couple of
equations of the form \eqref{eq:module}, one with $c_i(\cdot)$
monotone strictly increasing and the other with $c_i(\cdot)$
monotone strictly decreasing. We can therefore use the conditions
\eqref{eq:cond1} and \eqref{eq:cond2} with $M=2N$ for the analysis of
the Repressilator network, and can conclude that the equilibrium point of the Repressilator is LAS when
\begin{equation}
  \label{eq:lasrep}
  \frac{2\ptot \alpha \beta \frac{u^*}{\bbeta} }{\bar{\beta}K\gamma \left(1 + \frac{1}{K}(\frac{u^*}{\bbeta})^2\right)^2} < \sec\left(\frac{\pi}{2N}\right)^2,
\end{equation}
where $u^*$ is the unique solution to
\begin{align}
  \label{eq:f1}
  \frac{\alpha\beta P^\mrm{tot}}{\gamma\Big(1 + \frac{(\frac{u^*}{\bbeta})^2}{K}\Big)} = u^*,
\end{align}
and GAS when
\begin{equation}
  \label{eq:gasrep}
  \frac{3\ptot \alpha\beta}{8\overline{\beta}\gamma}\sqrt{\frac{3}{K}}< \sec\left(\frac{\pi}{2N}\right)^2.
\end{equation}
For simplicity, we assumed a Hill Coefficient $q=2$. We emphasize that
all the parameters that appear in equations
\eqref{eq:lasrep}--\eqref{eq:gasrep} are intrinsic to a single
repressor module, showing how we can again infer stability properties
of Repressilator from properties of a single TR repressor module. The
details of the computations that lead to
\eqref{eq:lasrep}--\eqref{eq:gasrep} can be found in Appendix~\ref{app:negfbk}.

\medskip

Figure~\ref{fig:results} shows the stability regions for a 3-gene
Repressilator as we vary two of the TR module parameters.

\section{Conclusions and Future Work}

We addressed the decomposition of biochemical networks into functional
modules, focusing our efforts on demonstrating how the behavior of a
complex network can be inferred from the properties of its constituent
modules.

\medskip

To illustrate our approach, we studied systems that have received
significant attention in the systems biology literature: gene
regulatory networks, enzymatic networks, and signaling pathways. Our
primary goal in defining these modules was to make sure that the
modules exhibited dynamic and parametric modularity, while making sure
that they had a clear biological function: regulating the production
of a protein, transforming a substrate into a different protein, or
transmitting information within a cell.

\medskip

A key issue that has not been addressed here is how one could go about
partitioning a biological network whose function is not known a priori, into a set of biologically ``meaningful'' modules. In the terminology
of Section~\ref{sec:probstat}, this question could be stated as how to
determine a partition for the dynamic DBSR graph that leads to
biologically ``meaningful'' modules --- recall that this section
provides a procedure to define the modules, \emph{assuming that a
  partition of the dynamic DBSR graph has been given.} We conjecture
that biologically meaningful partitions can be obtained by minimizing
the number of signals needed to interconnect the modules, which
corresponds to minimizing the number of arrows in the dynamic DBSR
graph that need to be ``severed'' by the partition
(cf.~Rule~\ref{ru:dbg}). The rationale for this conjecture lies in the
principle that functionally meaningful modules should be sparsely
connected, processing a small number of inputs to produce a small
number of outputs. In fact, most of the modules that we introduced in Section~\ref{sec:examples} have a single input and a single output, with the exception of the multi-gene TR module and the PD-cycle module that needs two inputs and two outputs for the bidirectional transmission of information in the signaling pathway.
\medskip

An important question in our current research is precisely to develop methods to partition large biological networks into biologically ``meaningful'' modules, using methods similar to those introduced in \cite{saez2008automatic,papachris}. This could allow us to ``discover'' biological function in these networks.

\section*{Acknowledgments}
The authors would like to acknowledge that the material presented in the paper is based upon work supported by the National Science Foundation under Grants No. ECCS-0835847 and EF-1137835.
\newpage
\section*{APPENDIX}
\appendix
\numberwithin{equation}{section} 
\numberwithin{figure}{section} 

\section{Module Characteristics}
\label{app:modchar}

In this section, we present some properties of the modules that could
not be fit into the main paper.

\subsection*{Transcriptional regulation (TR) module}

In Table \ref{tab:mod} of the main paper, we presented the LTF of a TR
module when Assumptions~\ref{as:tr-homogeneous}--\ref{as:tr-timescale}
were satisfied. A comprehensive explanation of how the dynamics of a
TR module are simplified with Assumption \ref{as:tr-timescale} is
provided in \cite{DelVecchio08}.

\medskip

It is reasonably straightforward to compute the LTF of a TR module
when only Assumption \ref{as:tr-homogeneous} is satisfied, and it is given by
\begin{align*}
H_j(s)= D_j\frac{q \kon\frac{K\ptot(\theta u^*)^q}{K+(\theta u^*)^q} \alpha_j \beta_j}{\Big(\theta u^*(s+\koff)(s+\bbeta) + \kon(\theta u^*)^q\big((q^2 F \frac{K\ptot(\theta u^*)^q}{K+(\theta u^*)^q} + \theta u^*)s + u^* \big)\Big)(s+\gamma_j)}
\end{align*}
with
\begin{align*}
D_j&\eqdef
\begin{cases}
+1 & \text{if $\ms_0$ activates $\gene{j}$}\\
-1 & \text{if $\ms_0$ represses $\gene{j}$},
\end{cases}
\qquad \theta \eqdef\frac{1}{\bar{\beta}}.
\end{align*}
This LTF now depends on the fan-out $F$. It is worth noting that the LTFs to each activating and to each repressing output are exactly the same because of Assumption~\ref{as:tr-homogeneous}. If this were not the case, the same method could be utilized to compute the LTF, but the expression would look different for each output. \cite{sivakumar2013towards} contains more information on how to compute the LTF of the module given the ODEs.

\subsection*{Enzyme-substrate reaction (ESR) module}
\label{sec:esrmchar}
The properties of ESR modules are summarized in Table \ref{tab:esrm}
of the main paper. The characteristics of these modules have been
presented using two classical approximations to simplify the reaction
dynamics.

\medskip

The form of the LTF obtained when using the quasi steady-state
approximation (Assumption~\ref{as:esr-quasi} from the main paper) was compact enough to fit in the main paper, but not the
LTF when using the equilibrium approximation (Assumption~\ref{as:esr-equilibrium} from the main paper). Under Assumption \ref{as:esr-equilibrium}, the LTF of an ESR module is given by

\begin{align*}
\frac{K^\mrm p(u^*)}{s+ K^\mrm q(u^*)}
\end{align*}
where
\begin{align*}
K^\mrm p(u^*) &= \frac{\etot \kcat K^\mrm d}{(K^\mrm d + \cX{\ms_0}^*)^2}\\
K^\mrm q(u^*) &= \frac{N}{\left(\etot K^\mrm d + (K^\mrm d + \cX{\ms_0}^*)^2\right)^2}\\
\intertext{with}
N &= \etot K^\mrm d\Big((K^\mrm d + \cX{\ms_0}^*)\big(\kcat(K^\mrm d+\cX{\ms_0}^*) + \\
&\qquad \gamma(K^\mrm d+3\cX{\ms_0}^*)-2u^*\big) + \kcat \etot(K^\mrm d + 2\cX{\ms_0}^*)\Big) + \gamma (K^\mrm d+\cX{\ms_0}^*)^4\\
\intertext{and}
\cX{\ms_0}^* &= \frac{-\etot \kcat + u^* - \gamma K^\mrm d + \sqrt{(-\etot \kcat + u^* - \gamma K^\mrm d)^2 + 4u^*\gamma K^\mrm d}}{2\gamma}.
\end{align*}

\subsection*{PD-cycle module}
The PD-cycle module was discussed in detail in
Section~\ref{subsec:pdcycle} of the main paper, but the LTF was omitted due to lack of
space and is shown in this section.

\medskip

The equilibrium point of the module as a function of the constant inputs $v_i^*$,
$u_i^*$ are given by
\begin{align*}
\cX{\ms_{i+1}}^* &= \frac{\kbarf_i(\kr_i +\alpha_i)v_i^*(\etot_i
	\albar_i - u_i^*)}{\kf_i \alpha_i (\kbarr_i + \albar_i)u_i^*}, &
[\bndxy{\ms_i^\dagger}{\ms_{i+1}}]^* &= \frac{v_i^*}{\alpha_i}\\
[\ms_i^\dagger]^* &= \frac{u_i^*(\kbarr_i
	+\albar_i)}{\kbarf_i(\etot_i \albar_i - u_i^*)}, &
\cX{E_{i}}^* &= \etot_i -\frac{u_i^*}{\albar_i}
\end{align*}
and the LTF around this equilibrium is given by
\begin{equation*}
H(s) 
\eqdef \frac{1}{D(s)}\begin{bmatrix} N_{11}(s) & N_{12}(s) \\ N_{21}(s) & N_{22}(s)\end{bmatrix}
\end{equation*}
where
\begin{align*}
N_{11}(s) & =  {\kf_i} {[\ms_i^\dagger]^*} ({\cX{E_{i}}^*} {\kbarf_i} (s+{\albar_i})+s ({\kbarr_i}+s+{\kbarf_i} {[\ms_i^\dagger]^*}+{\albar_i}))
{\alpha_i} \\
N_{12}(s) &=  {\kf_i} s {\cX{\ms_{i+1}}^*} ({\kbarr_i}+s+{\kbarf_i} {[\ms_i^\dagger]^*}+{\albar_i}) {\alpha_i}\\
N_{21}(s) &=  -{\cX{E_{i}}^*} {\kbarf_i} {\kf_i} s {[\ms_i^\dagger]^*} {\albar_i}\\
N_{22}(s) &= {\cX{E_{i}}^*} {\kbarf_i} {\albar_i} ({\kr_i} s+(s+{\kf_i}
{[\ms_i^\dagger]^*}) (s+{\alpha_i})),\\
D(s) &= \cX{E_i}^*\kbarf_i(s+\albar_i)\big(\kr_i s + (s + \kf_i \cX{\ms_i}^*)(s+\alpha_i)\big) + \\
&\quad  s\big(s + \kbarr_i + \kbarf_i\cX{\ms_i}^* +\albar_i\big)\big(s^2 + (\kr_i + \kf_i([\bndxy{\ms_i^\dagger}{\ms_{i+1}}]^* + \cX{\ms_i}^*)+\alpha_i)s +\alpha_i\kf_i\cX{\ms_i}^*\big).
\end{align*}
\newpage
\section{Covalent Modification Network}
\label{app:posfbk}

To prove the statements from the main paper about the Covalent Modification network, we use the following result which is adapted from \cite[Theorems 2-3]{angeli2004multi}, and provides
conditions that can be used to establish the stability of the
equilibrium points for positive feedback networks. To express the
result, we need the following definitions which are closely related to that of
cooperativity: We say that the system
\begin{align}\label{eq:nonlinear-1}
\dot x&=A(x,u), &
y&= B(x), & 
x\in\R^n,u\in\R^k,y\in\R^m,
\end{align}
is \emph{excitable (with respect to the positive orthant)} if for
every initial condition $x_0\in\R^n$ and all inputs $u(t), \bar
u(t)\in\R^k$, $\forall t\ge 0$, we have that
\begin{align*}
u(t)\succ \bar u(t),\;\forall t\ge 0 
\imply 
x(t;x_0,u)\gg x(t;x_0,\bar u) ,\;\forall t> 0, 
\end{align*}
and it is \emph{transparent (with respect to the positive orthant)} if
for all initial conditions $x_0,\bar x_0\in\R^n$ and every input
$u(t)\in\R^k$, $\forall t\ge 0$, we have that
\begin{align*}
x_0\succ \bar x_0 
\imply 
x(t;x_0,u)\gg x(t;\bar x_0, u) ,\;\forall t> 0.
\end{align*}
Given two vectors $v,\bar v$, we write $v\succ \bar v$ if every entry
of $v$ is larger than or equal to the corresponding entry of $\bar v$
and $v\neq \bar v$.
\begin{theorem}\label{thm:monotone}
	Consider the feedback interconnection depicted in
	Figure~\ref{fig:blockfeed} with $u(t)=0$, $\forall t\ge 0$ and
	$\mv_1$ SISO of the form \eqref{eq:nonlinear-1}. Assume that
	\begin{enumerate}
		\item \label{subthm:ex}$\mv_1$ is excitable, transparent, and cooperative with a well-defined ISSCF and IOSCF;
		\item \label{subthm:gas} for every constant input $u_1(t)=u_1^*$, $\forall t\ge 0$ to
		$\mv_1$, the Jacobian matrix $\PDeriv{A(x,u_1)}{x}$ is nonsingular
		at the corresponding equilibrium, which is globally asymptotically
		stable;
		\item \label{subthm:ioscf} the IOSCF $f_1(u_1^*)$ of $\mv_1$ has fixed points $u^*$ for
		which $f_1(u^*)=u^*$ and $\PDeriv{f_1(u_1^*)}{u_1^*}\Big|_{u_1^* = u^*}\neq 1$;
		\item \label{subthm:bounded} all trajectories of the feedback interconnection are bounded.
	\end{enumerate}
	Then, for almost all initial conditions, the solutions converge to
	the set of points for which $f_1(u_1^*)=u_1^*$ and
	$\PDeriv{f_1(u_1^*)}{u_1^*}~<~1$.\frqed
\end{theorem}
We now use Theorem \ref{thm:monotone} to prove Corollary \ref{cor:covmod}.

\begin{corollary}\label{cor:covmod}
	Consider the covalent modificiation network shown in Figure
	\ref{fig:covmodi}, which consists of a cascade of two ESR modules
	connected in positive feedback. The substrate concentrations in each
	module converges to $0$ as $t \to \infty$. \frqed
\end{corollary}
\begin{proof-corollary}{\ref{cor:covmod}}
	We consider the covalent modificiation network adapted from
	\cite{goldbeter1981amplified}, which is a
	network given by the chemical reactions
	\begin{align}
	\label{eq:network}
	\begin{split}
	\orxna{\ms_1 + E_1 \xrightleftharpoons[k^{\textup{r}}_1\cX{\bndxy{\ms_1}{E_1}}]{k^{\textup{f}}_1\cX{\ms_1}\cX{E_1}}\bndxy{\ms_1}{E_1}}{k^{\tu{cat}}_1\cX{\bndxy{\ms_1}{E_1}}}{\ms_2 + E_1}\\
	\orxna{\ms_2 + E_2 \xrightleftharpoons[k^{\textup{r}}_2\cX{\bndxy{\ms_2}{E_2}}]{k^{\textup{f}}_2\cX{\ms_2}\cX{E_2}}\bndxy{\ms_2}{E_2}}{k^{\tu{cat}}_2\cX{\bndxy{\ms_2}{E_2}}}{\ms_1 + E_2}\\
	\orxna{\ms_1}{\gamma_1\cX{\ms_1}}{\varnothing}\\
	\orxna{\ms_2}{\gamma_2\cX{\ms_2}}{\varnothing}.
	\end{split}
	\end{align}
	This network can be viewed as a positive feedback interconnection of a
	cascade of two ESR modules described in Table \ref{tab:esrm} of the
	main paper, one containing the substrate $\ms_1$ and the other the
	substrate $\ms_2$. We call these modules $\mv_1$ and
	$\mv_2$, and the module dynamics for $i =\{1,2\}$ are each given by
	\begin{align*}
	\bm{\mv_i}: \dotcX{\ms_i} &=  A_i(\cX{\ms_i},u_i), &
	y_i &= B_i(\cX{\ms_i})
	\end{align*}
	where
	\begin{align}
	\label{eq:eqapprox}
	A_i(\cX{\ms_i},u_i) &= \frac{u_i - \gamma_i \cX{\ms_i} -
		\frac{\kcat_i \etot_i\cX{\ms_i}}{\Kd_i + \cX{\ms_i}}}{1 +
		\frac{\Kd_i \etot_i}{(\Kd_i +\cX{\ms_i}^2)}}  &
	B_i(\cX{\ms_i}) &= \frac{k^\mrm{cat}\etot_i\cX{\ms_i}}{\Kd_i + \cX{\ms_i}}
	\end{align}
	under Assumption
	\ref{as:esr-equilibrium}, and 
	\begin{align}
	\label{eq:qssapprox}
	A_i(\cX{\ms_i},u_i)  &= u_i - \gamma_i \cX{\ms_i} -
	\frac{k^\mrm{cat}_iE^{\tu{tot}}_i\cX{\ms_i}}{K^\mrm{m}_i + \cX{\ms_i}} &
	B_i(\cX{\ms_i}) &= \frac{k^\mrm{cat}_iE^{\tu{tot}}_i\cX{\ms_i}}{K^\mrm{m}_i + \cX{\ms_i}}
	\end{align}
	under Assumption
	\ref{as:esr-quasi}. The covalent modification
	system consists of the modules $\mv_1$ and $\mv_2$ interconnected with
	$y_1 = u_2$ and $y_2 = u_1$.
	
	\medskip
	
	We first need to show that the cascade of two ESR modules satisfies the
	properties in item \ref{subthm:ex} from Theorem
	\ref{thm:monotone}. For the cascade network with $y_1 = u_2$ and the
	module dynamics \eqref{eq:eqapprox} or \eqref{eq:qssapprox}, we
	have
	\begin{align*}
	\PDeriv{A_1(\cX{\ms_1},u_1)}{u_1} = 1 &  \qquad \PDeriv{A_2(\cX{\ms_2},u_2)}{\cX{\ms_1}} >0 \qquad \PDeriv{y_2}{\cX{\ms_2}} > 0 \\ 
	\PDeriv{A_1(\cX{\ms_1},u_1)}{\cX{\ms_2}} = 0 &\qquad \PDeriv{A_2(\cX{\ms_2},u_2)}{u_1} = 0 \qquad \PDeriv{y_2}{\cX{\ms_1}} = 0,
	\end{align*}
	from which we can infer several properties: From \cite[Propositon
	1]{angeli2004interconnections}, we conclude that the cascade is
	cooperative, from \cite[Theorem 4]{angeli2004multi} that it is
	excitable, and from \cite[Theorem 5]{angeli2004multi} that it is
	transparent. For simplicity, in the rest of the proof we use
	Assumption~\ref{as:esr-equilibrium} and therefore the module dynamics
	in \eqref{eq:eqapprox}, but the proof would be similar for the module
	dynamics in \eqref{eq:qssapprox}.
	
	\medskip
	The IOSCF of the
	cascade is well-defined and given by $f_2(f_1(u_1^*))$, where $u_1^*$
	denotes a constant input to the cascade and 
	\begin{align*}
	f_i(u_i^*) &\eqdef y_i^* = \frac{1}{2}(u_i^* + k^\mrm{cat}_i E^{\tu{tot}}_i + \Kd_i \gamma_i -\sqrt{(-u_i^* + k^\mrm{cat}_i E^{\tu{tot}}_i + \Kd_i \gamma_i)^2 + 4\Kd_i u_i^* \gamma_i}) \qquad i = \{1,2\}.
	\end{align*}
	The ISSCF of the cascade is also well-defined, with the equilibrium
	states corresponding to a constant input $u_1^*$ to the cascade given by
	\begin{align*}
	\cX{\ms_1}^* = g_1(u_1^*),\quad
	\cX{\ms_2}^* = g_2(g_1(u_1^*))
	\end{align*}
	where
	\begin{align*}
	g_i(u_i^*) &= \frac{1}{2}(K^\mrm{p}_i + \sqrt{4K^\mrm{d}_i\gamma_i u_i^* + (K^\mrm{p}_i)^2})&
	K^\mrm{p}_i &\eqdef u_i^* -K^\mrm{d}_i\gamma_i -  k^\mrm{cat}_iE^{\tu{tot}}_i,&
	i\in\{1,2\}.
	\end{align*}
	Therefore the cascade of two ESR modules satisfies the properties in
	item \ref{subthm:ex} of Theorem \ref{thm:monotone}.
	
	\medskip
	
	We now show that the cascade of two ESR modules satisfies the
	properties in item \ref{subthm:gas} from Theorem~\ref{thm:monotone}. For some constant input $u_1^* \geq 0$, the
	Jacobian matrix of the interconnection is given by the $2\times 2$ lower
	triangular matrix 
	\begin{align*}
	J = \begin{bmatrix}
	J_{11} & 0 \\ J_{21} & J_{22},
	\end{bmatrix}
	\end{align*}
	which is non-singular because $J_{11}>0$ and $J_{22}>0$, under the
	implicit assumption that all parameters within each module are
	positive.
	
	\medskip
	
	Each of the modules $\mv_1$ and $\mv_2$ has an equilibrium point that
	is globally asymptotically stable for some constant input $u_i^*$ into
	each module. This can be verified by doing the coordinate
	transformation $x_i = \cX{\ms_i} - \cX{\ms_i}^*$ and observing that
	the Lyapunov function $V(x_i) = x_i^2$ is zero-at-zero, locally
	positive definite and $\dot V(x_i)<0$, $\forall x_i$, $i \in
	\{1,2\}$. The cascade of both the modules also has a globally
	asymptotically stable equilibrium point, as can be seen from the
	argument in \cite{sundarapandian2002global}.
	
	\medskip
	
	To verify that the cascade of the ESR modules satisfies the property
	in item \ref{subthm:ioscf} from Theorem \ref{thm:monotone}, we will show
	that $f_2(f_1(u_1^*)) = u_1^*$ has a unique solution at
	$u_1^* =0$, and also that
	\begin{align*}
	\PDeriv{f_2(f_1(u_1^*))}{u_1^*}\Bigg|_{u_1^* = 0} <1.
	\end{align*}
	To show that $f_2(f_1(u_1^*)) = u_1^*$ has a unique solution
	at $u_1^* =0$, we first observe that the IOSCFs of $\mv_1$ and
	$\mv_2$ are each monotone increasing and strictly concave, since
	$f_i'(u_i^*)>0$ and $f_i''(u_i^*) <0$, $\forall i \in \{1,2\}$. Then from
	\begin{align}
	\label{eq:1}
	\frac{\partial f_2\left(f_1\left(u_1^*\right)\right)}{\partial u_1^*} &= f_1'(u_1^*)f_2'\left(f_1\left(u_1^*\right)\right)\\
	\label{eq:2}
	\frac{\partial^2 f_2\left(f_1\left(u_1^*\right)\right)}{\partial (u_1^*)^2} &= f_1''(u_1^*)f_2'\left(f_1\left(u_1^*\right)\right) + \left(f_1'\left(u_1^*\right)\right)^2 f_2''\left(f_1\left(u_1^*\right)\right),
	\end{align}
	it can be seen that the cascade of $\mv_1$ and $\mv_2$ is also monotone increasing and strictly concave because
	\begin{enumerate}
		\item $f_1'(u_1^*) >0$ and $f_2'(u_2^*) >0 \quad \forall u_1^*,u_2^*>0 \implies \frac{\partial f_2\left(f_1\left(u_1^*\right)\right)}{\partial u_1^*} >0$ from \eqref{eq:1}.
		\item$f_1''(u_1^*) <0$ and $f_2''(u_2^*) < 0 \quad \forall u_1^*,u_2^*>0 \implies \frac{\partial^2 f_2\left(f_1\left(u_1^*\right)\right)}{\partial (u_1^*)^2} < 0$ from \eqref{eq:2}.
	\end{enumerate}
	Therefore, there can be no other solution to $f_2(f_1(u_1^*)) =u_1^*$ other than $u_1^* =0$. In addition, the IOSCF of the
	cascade is given by
	\begin{align*}
	\label{eq:covmodder}
	\PDeriv{f_2(f_1(u_1^*))}{u_1^*}\Big|_{u_1^* = 0} = \frac{\etot_1\etot_2\kcat_1\kcat_2}{(\etot_1\kcat_1 + K^d_1 \gamma_1)(\etot_2\kcat_2 + K^d_2\gamma_2)},
	\end{align*}
	which is always less than $1$ since all parameters are positive by
	definition.
	
	\medskip
	
	The boundedness property in item \ref{subthm:bounded} of Theorem
	\ref{thm:monotone} follows from techniques used to analyze MAK ODEs from \cite[Main Technical Lemma]{angeli2011boundedness}, which can be applied to the covalent modificiation network.
	
\end{proof-corollary}
\newpage
\section{Repressilator network}
\label{app:negfbk}
\subsection*{Generalized Nyquist Criterion}
\begin{figure}[h]
	\centering
	\psfrag{M1}[b][b]{\Large $\mv_1$}
	\psfrag{M2}[b][b]{\Large $\mv_2$}
	\psfrag{M3}[b][b]{\Large $\mv_3$}
	\psfrag{MN}[b][b]{\Large $\mv_N$}
	\psfrag{H1}[][]{$H_1(s)$}
	\psfrag{u1}[r][r]{\scriptsize$u_1$}
	\psfrag{u2}[][]{\scriptsize$u_2$}
	\psfrag{u3}[][]{\scriptsize$u_3$}
	\psfrag{uN}[r][r]{\scriptsize$u_N$}
	\psfrag{y1}[][]{\scriptsize$y_1$}
	\psfrag{y2}[][]{\scriptsize$y_2$}
	\psfrag{y3}[][]{\scriptsize$y_3$}
	\psfrag{yN}[l][l]{\scriptsize$y_N$}
	\psfrag{u}[r][l]{\scriptsize$u$}
	\psfrag{y}[l][r]{\scriptsize$y$}
	\includegraphics[width = 0.9\textwidth]{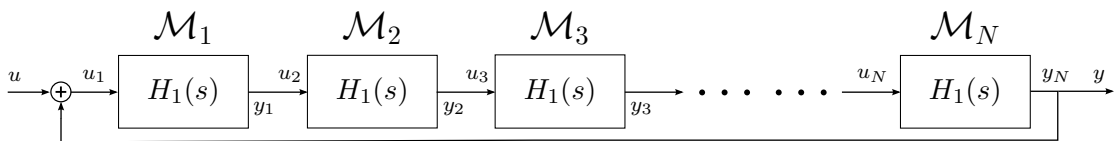}
	\caption{Cascade of $N$ modules, each with an
		equal LTF $H_1(s)$, connected in feedback. Theorem \ref{thm:nyq2}
		provides conditions for the BIBO stability of the linearized
		network, from a small perturbation in $u$ to a small perturbation
		in $y$. }
	\label{fig:fbk_gen}
\end{figure}

We first present a result that will be used to prove Corollary \ref{cor:necsuff}. The following result is inspired by the Nyquist Stability Criterion \cite{dorfbook}
and provides a necessary and sufficient condition to establish the
BIBO stability of the LTF of a feedback interconnection of
a cascade of $N\ge 1$ modules, each with an equal LTF as depicted in
Figure~\ref{fig:fbk_gen}.
\begin{theorem}
	\label{thm:nyq2}
	Consider the feedback interconnection of a cascade of $N$ modules
	depicted in Figure~\ref{fig:fbk_gen}, all with the same LTF
	$H_1(s)$ around a given equilibrium point of the feedback
	interconnection. Then the LTF of the feedback connection is BIBO
	stable if and only if
	\begin{equation*}
	\#OUP=-\frac{1}{N}\sum_{\ell=1}^{N} \#END[e^{j \frac{2\pi \ell}{N}}],
	\end{equation*}
	where $\#OUP$ represents the number of (open-loop) unstable poles of
	$H_1(s)$ and $\#END[e^{j \frac{2\pi \ell}{N}}]$ denotes the number
	of clockwise encirclements of the Nyquist contour of $H_1(j
	\omega)$, $\omega\in\R$ around the point $e^{j \frac{2\pi\ell}{N}}$
	on the complex plane.\footnote{We assume here that $H_1(s)$ has no
		poles on the imaginary axis. If this were the case, the standard
		``trick'' of considering an infinitesimally perturbed system with the
		poles moved off the axis can be applied \cite{dorfbook}.}\frqed
\end{theorem}

\begin{proof-theorem}{\ref{thm:nyq2}}
	To investigate the BIBO stability of the LTF of the network in Figure \ref{fig:fbk_gen}, we consider the characteristic equation
	of the feedback loop: $1 - H_1(s)^N = 0$. The number of
	unstable poles is thus given by the unstable solutions to the
	equation:
	\begin{equation*}
	1 - H_1(s)^N = 0
	\eqv \exists i\in\{1,2,\cdots,n\}, \; H_1(s) = z_i,
	\end{equation*}
	where $z_{\ell}\eqdef e^{j \frac{2\pi \ell}{N}}$ are the $N$ roots
	to the equation $z^N=1$.
	
	\medskip
	
	To count the number of unstable poles of the network, we must then
	add the number of unstable poles of each of the $N$ equations
	\begin{equation*}
	H_1(s) = z_\ell, \quad \ell\in\{1,2,\cdots,N\},
	\end{equation*}
	which can be done using Cauchy's argument principle by counting the
	number of clockwise encirclements of the point $z_\ell\in\C$ for the Nyquist
	contour of $H_1(j\omega)$, $\omega\in\R$. \frQED
\end{proof-theorem}

\begin{corollary}
	\label{cor:necsuff}
	Consider a Repressilator network, that consists of an odd number $N$
	of equal single-gene TR repressor modules ($F=1$) connected in
	feedback as in Figure~\ref{fig:fbk_spc}. The
	network has a unique equilibrium point that is locally
	asymptotically stable if and only if
	\begin{align*}
	\sum_{\ell=1}^{N} \#END[e^{j \frac{2\pi \ell}{N}}]=0,
	\end{align*}
	where $\#END[e^{j \frac{2\pi \ell}{N}}]$ denotes the number of
	clockwise encirclements of the Nyquist plot of the LTF of a single TR repressor module around the point $e^{j \frac{2\pi\ell}{N}}$. \frqed
\end{corollary}
\begin{proof-corollary}{\ref{cor:necsuff}}
	We assume that Assumption \ref{as:tr-timescale} is satisfied for simplicity, although it is straightforward to extend the proof for the case when it is not. Corollary \ref{cor:necsuff} follows from Theorem~\ref{thm:nyq2} by
	recognizing that a single-gene repressor TR module can be
	represented by the following LTF
	\begin{align}
	\label{eq:tr-h1}
	H_1(s)&= -\frac{q K  P^\mrm{tot}\alpha_1\beta_1 (\frac{u^*}{\bbeta})^{q-1}}{(K +(\frac{u^*}{\bbeta})^q)^2 (s+\gamma_1)(s+\bar{\beta})}
	\end{align}
	under Assumption \ref{as:tr-timescale}. Moreover, since the network input $u(t) =0$ $\forall t\geq 0$ and all modules have identical parameters, the values of the inputs and outputs of each module at equilibrium will be the same, each given by the unique solution $u^*$ to \eqref{eq:f1}. Therefore all modules have
	equal LTFs, and this enables us to use Theorem~\ref{thm:nyq2} to analyze when the LTF of the Repressilator is BIBO stable.
	
	\medskip
	
	To complete the proof, we need to show that if the LTF of the
	Repressilator network is BIBO stable, then the equilibrium point is
	locally asymptotically stable. To do this, we need to show that the
	realization of the linearized network is minimal
	\cite{Hespanha09}. Defining the state of the network to be
	\begin{align*}
	x = \begin{bmatrix}  \cX{\ms_1} \\\cX{mRNA_1}\\\cX{\ms_2} \\  \cX{mRNA_2}\\
	\vdots \\ \vdots \\ \cX{\ms_N}\\ \cX{mRNA_N}
	\end{bmatrix},
	\end{align*}
	the state-space realization of the linearized closed-loop
	Repressilator network is given by
	\begin{align*}
	\dot {\delta x} = A\delta x + B\delta u \qquad \delta y = C \delta x
	\end{align*}
	where
	\begin{align*}
	A_{2N\times 2N} &= \begin{bmatrix}
	-\bbeta & 0 & 0 & 0 & 0 & \cdots &0&0&0& 0 & \beta\\ P & -\gamma & 0 & 0 & 0 &\cdots & 0 &0&0&0& 0 \\0 & \beta & -\bbeta & 0 & 0 &\cdots & 0 & 0&0&0&0\\ 0 & 0 & P & -\gamma & 0 &\cdots & 0&0&0&0 & 0\\ \vdots & \vdots &\vdots &\vdots &\vdots & \ddots &\vdots & \vdots&\vdots &\vdots&\vdots \\ \vdots & \vdots & \vdots & \vdots &\vdots & & \ddots & \vdots&\vdots &\vdots&\vdots\\ 0 & 0 & 0 & 0 & 0 &\cdots & \beta&-\bbeta&0&0 & 0\\ 0 & 0 & 0 & 0 & 0 &\cdots & 0&P&-\gamma&0 & 0\\ 0 & 0 & 0 & 0 & 0 &\cdots & 0&0&\beta&-\bbeta & 0\\ 0 & 0 & 0 & 0 & 0 &\cdots & 0&0&0&P &-\gamma
	\end{bmatrix}\\
	B_{2N\times 1} &= \begin{bmatrix}1 \\0\\ \vdots\\0\\0\end{bmatrix}\qquad C_{1\times 2N} = \begin{bmatrix}0&0&\cdots&0&\bbeta\end{bmatrix}
	\qquad 
	P = -\frac{q\alpha\ptot K (\cX{\ms_i}^*)^{q-1} }{(K + (\cX{\ms_i}^*)^{q})^2} \quad \forall i.
	\end{align*}
	Both the controllability and observability matrices of this system
	have full rank and therefore the realization is minimal
	\cite{Hespanha09}. In this case, BIBO stability of the LTF implies
	that the realization is exponentially stable and therefore the
	equilibrium is locally asymptotically stable. \frqed
	
\end{proof-corollary}
Figure \ref{fig:nyquist1} shows a numerical example of the Nyquist plot of \eqref{eq:tr-h1} when $N=3$. When Assumption~\ref{as:tr-timescale} is not satisfied, \eqref{eq:tr-h1} is given by the more complicated LTF that is presented in Appendix~\ref{app:modchar}. 

\draftbreak 
\subsection*{Generalized Secant Criterion}
The following result provides a alternative condition that can
be used to establish the stability of a negative feedback
interconnection of a cascade of $N$ modules, not necessarily with
the same LTFs. This result will be used to prove Corollary \ref{cor:suff}, but can be used for a wide variety of networks.
This new condition is inspired by results in
\cite{tyson1978dynamics,thron1991secant,arcak2006diagonal} and applies
to the feedback interconnection depicted in Figure \ref{fig:fbk_spc}, where the modules $\mv_i$ are
each described by equations of the following form
\begin{align}
\label{eq:module}
\dot{\cX{\ms_i}} &= u_i - c_i(\cX{\ms_i}), & y_i &= d_i(\cX{\ms_i}),
& i\in\{1,2,\dots,N\},
\end{align}
for appropriate scalar functions $c_i(\cdot)$, $d_i(\cdot)$. All the
functions $c_i(\cdot)$ are assumed to be monotone strictly increasing
and the functions $d_i(.)$ are assumed to be either monotone strictly
decreasing or strictly increasing. The IOSCF of the $i$th module is
given by
\begin{align*}
f_i(u_i^*)=d_i\big(c_i^{-1}(u_i^*)\big),\quad\forall u_i^*\in\R,
\end{align*}
where $c_i^{-1}(\cdot)$ denotes the inverse function of $c_i(\cdot)$,
which is invertible and monotone strictly increasing since
$c_i(\cdot)$ is monotone strictly increasing. Therefore, $f_i(u_i^*)$
has the same (strict) monotonicity as $d_i(\cdot)$. The result that
follows considers the case in which we have an odd number of
$d_i(\cdot)$ that are monotone strictly decreasing and the remaining
are monotone strictly increasing. In this case, the composition of the
IOSCFs of the $N$ modules is monotone strictly decreasing and we have
a negative feedback interconnection.

\begin{theorem}
	\label{thm:secdiag}
	Consider the feedback interconnection depicted in
	Figure~\ref{fig:fbk_spc}, with each of the $N$ modules $\mv_i$ of
	the form \eqref{eq:module}, where
	\begin{enumerate}
		\item the $c_i(\cdot)$ are continuous and monotone strictly increasing;
		\item an odd number $M\le N$ of the $d_i(\cdot)$ are continuous and monotone
		strictly decreasing, while the remaining $N-M$ of the $d_i(\cdot)$
		are continuous and monotone strictly increasing.
	\end{enumerate}
	Then the IOSCF of the cascade is monotone strictly decreasing and
	the feedback interconnection has a unique equilibrium. This
	equilibrium is locally asymptotically stable provided that
	\begin{align}
	\label{eq:cond1}
	\prod_{i=1}^{N} \bigg|\frac{\PDeriv{d_i(s_i)}{s_i}\big|_{s_i=\cX{\ms_i}^*}}{\PDeriv{c_i(s_i)}{s_i}\big|_{s_i=\cX{\ms_i}^*}}\bigg|
	< \sec\left({\frac{\pi}{N}}\right)^N,
	\end{align}
	where $\cX{\ms_i}^*$ denotes the value of $\cX{\ms_i}$ at the
	equilibrium; and it is globally asymptotically stable if there
	exist constants $\phi_i>0$, $i\in\{1,2,\dots,N\}$ for which
	\begin{align*}
	\Big|\frac{d_i(z_i) - d_i(\cX{\ms_i}^*)}{c_i(z_i) - c_i(\cX{\ms_i}^*)} \Big| \leq \phi_i \quad \forall z_i \neq \cX{\ms_i}^*
	\end{align*}
	and
	\begin{align*}
	\prod_{i=1}^{N}\phi_i &< \sec\left({\frac{\pi}{N}}\right)^N.\tag*\frqed
	\end{align*}
\end{theorem}
An alternative condition to \eqref{eq:gas} which is simpler to verify is given by
\begin{align*}
\Big|\PDeriv{d_i(z_i)}{z_i}\Big| \leq \phi_i\PDeriv{c_i(z_i)}{z_i},\quad \forall z_i \neq \cX{\ms_i}^*.
\end{align*}

\begin{proof-theorem}{\ref{thm:secdiag}}
	The proof of this result relies on making a coordinate
	transformation to our original system to take it into a form that
	allows us to use the secant criterion
	\cite{tyson1978dynamics,thron1991secant,arcak2006diagonal} and
	results in \cite{arcak2006diagonal}.
	
	\medskip
	
	The dynamics of the feedback interconnection under consideration can
	be written as
	\begin{subequations}\label{eq:dynamics}
		\begin{align}
		\dot{\cX{\ms_1}}&=-c_1(\cX{\ms_1})+d_N(\cX{\ms_N}), \\
		\dot{\cX{\ms_2}}&=-c_2(\cX{\ms_2})+d_1(\cX{\ms_1}) \\
		&\;\vdots\\
		\dot{\cX{\ms_N}}&=-c_N(\cX{\ms_N})+d_{N-1}(\cX{\ms_{N-1}}).
		\end{align}
	\end{subequations}
	with an equilibrium state defined by concentrations $\cX{\ms_i}^*$ for which
	\begin{subequations}\label{eq:equilibrium}
		\begin{align}
		c_1(\cX{\ms_1}^*)&=d_N(\cX{\ms_N}^*), \\
		c_2(\cX{\ms_2}^*)&=d_1(\cX{\ms_1}^*), \\
		&\;\vdots\\
		c_N(\cX{\ms_N}^*)&=d_{N-1}(\cX{\ms_{N-1}}^*).
		\end{align}
	\end{subequations}
	To verify that such an equilibrium exists and is unique, note that we
	have a feedback interconnection of a cascade of $N$ systems, each with an
	IOSCF given by
	\begin{align*}
	f_i(u_i^*)=d_i\big(c_i^{-1}(u_i^*)\big),\quad\forall u_1^*\in\R,
	\end{align*}
	where $c_i^{-1}$ denotes the inverse function of $c_i$, which is
	invertible and monotone strictly increasing since $c_i$ is monotone
	strictly increasing. Therefore, $f_i(u_i^*)$ has the same (strict)
	monotonicity as $d_i$. Since an odd number $M$ of the $d_i$ are
	monotone strictly decreasing, an odd number of the $f_i$ are also
	monotone strictly decreasing and therefore the composition of
	all the $f_i$ is monotone strictly decreasing. This shows that we
	have a negative feedback interconnection and thus a unique
	equilibrium.
	
	\medskip
	
	Using \eqref{eq:equilibrium}, we can re-write \eqref{eq:dynamics} as
	\begin{align*}
	\dot{\cX{\ms_1}}&=-c_1(\cX{\ms_1})+c_1(\cX{\ms_1}^*)+d_N(\cX{\ms_N})-d_N(\cX{\ms_N}^*), \\
	\dot{\cX{\ms_2}}&=-c_2(\cX{\ms_2})+c_2(\cX{\ms_2}^*)+d_1(\cX{\ms_1})-d_1(\cX{\ms_1}^*), \\
	&\;\vdots\\
	\dot{\cX{\ms_N}}&=-c_N(\cX{\ms_N})+c_N(\cX{\ms_N}^*)+d_{N-1}(\cX{\ms_{N-1}})-d_{N-1}(\cX{\ms_{N-1}}^*).
	\end{align*}
	Since we have an odd number $M\ge 1$ of functions $d_i$ that
	are monotone strictly decreasing and there is perfect symmetry in
	the cycle \eqref{eq:dynamics}, we shall assume without loss of
	generality that $d_N$ is monotone strictly decreasing; if
	that were not the case we could simply shift the numbering of the
	modules appropriately.
	
	\medskip
	
	We consider a coordinate transformation with
	\begin{align}\label{eq:1-N}
	x_1&=\cX{\ms_1}-\cX{\ms_1}^*, &
	x_N&=\cX{\ms_N}-\cX{\ms_N}^*, 
	\end{align}
	and the remaining $x_i$, $i\in\{2,3,\dots,N-1\}$ either given by
	\begin{align}
	\label{eq:plus}
	x_i=\cX{\ms_i}-\cX{\ms_i}^*
	\intertext{or given by}
	\label{eq:minus}
	x_i=-\cX{\ms_i}+\cX{\ms_i}^*;
	\end{align}
	each to be determined shortly. The coordinate
	transformation \eqref{eq:1-N} leads to
	\begin{align*}
	\dot x_1&=-c_1(\cX{\ms_1})+c_1(\cX{\ms_1}^*)+d_N(\cX{\ms_N})-d_N(\cX{\ms_N}^*)\\
	&=-c_1(\cX{\ms_1}^*+x_1)+c_1(\cX{\ms_1}^*)+d_N(\cX{\ms_N}^*+x_N)-d_N(\cX{\ms_N}^*)\\
	&=-a_1(x_1)-b_N(x_N)
	\end{align*}
	with
	\begin{align*}
	a_1(x_1)&\eqdef c_1(\cX{\ms_1}^*+x_1)-c_1(\cX{\ms_1}^*), &
	b_N(x_N)&\eqdef -d_N(\cX{\ms_N}^*+x_N)+d_N(\cX{\ms_N}^*).
	\end{align*}
	Note that because $c_1$ is monotone strictly increasing and $d_N$ is
	monotone strictly decreasing, we have that
	\begin{align}\label{eq:ab-1-N}
	&a_1(x_1)\quad\begin{cases}
	>0 & x_1>0\\
	= 0 & x_1 = 0\\
	<0 & x_1<0.
	\end{cases}, &
	&b_N(x_N)\quad\begin{cases}
	>0 & x_N>0\\
	=0 & x_N = 0\\
	<0 & x_N<0.
	\end{cases}
	\end{align}
	For the remaining  variables $x_i$, $i\in\{2,3,\dots,N\}$, the coordinate transformation leads to
	\begin{align*}
	\dot x_i
	&=\\ &\begin{cases}
	c_i(\cX{\ms_i}^*)-c_i(\cX{\ms_i})+d_{i-1}(\cX{\ms_{i-1}})-d_{i-1}(\cX{\ms_{i-1}}^*)
	& \text{if }x_i=\cX{\ms_i}-\cX{\ms_i}^*,\:x_{i-1}=\cX{\ms_{i-1}}-\cX{\ms_{i-1}}^*\\
	& \text{or }x_i=\cX{\ms_i}-\cX{\ms_i}^*,\:x_{i-1}=\cX{\ms_{i-1}}^*-\cX{\ms_{i-1}}\\
	c_i(\cX{\ms_i})-c_i(\cX{\ms_i}^*)-d_{i-1}(\cX{\ms_{i-1}})+d_{i-1}(\cX{\ms_{i-1}}^*)
	& \text{if }x_i=\cX{\ms_i}^*-\cX{\ms_i},\:x_{i-1}=\cX{\ms_{i-1}}-\cX{\ms_{i-1}}^*\\
	& \text{or }x_i=\cX{\ms_i}^*-\cX{\ms_i},\:x_{i-1}=\cX{\ms_{i-1}}^*-\cX{\ms_{i-1}}
	\end{cases}\\
	&=-a_i(x_i)+b_{i-1}(x_{i-1})
	\end{align*}
	where, for every $i\in\{2,3,\dots,N\}$,
	\begin{align*}
	a_i(x_i) &\eqdef \begin{cases}
	c_i(\cX{\ms_i}^*+x_i)-c_i(\cX{\ms_i}^*) & x_i=\cX{\ms_i}-\cX{\ms_i}^*\\
	-c_i(\cX{\ms_i}^*-x_i)+c_i(\cX{\ms_i}^*) & x_i=\cX{\ms_i}^*-\cX{\ms_i}
	\end{cases}\\
	b_{i-1}(x_{i-1}) &\eqdef\\ &\begin{cases}
	d_{i-1}(\cX{\ms_{i-1}}^*+x_{i-1})-d_{i-1}(\cX{\ms_{i-1}}^*)
	& \text{if }x_i=\cX{\ms_i}-\cX{\ms_i}^*,\;x_{i-1}=\cX{\ms_{i-1}}-\cX{\ms_{i-1}}^*\\
	d_{i-1}(\cX{\ms_{i-1}}^*-x_{i-1})-d_{i-1}(\cX{\ms_{i-1}}^*)
	& \text{if }x_i=\cX{\ms_i}-\cX{\ms_i}^*,\;x_{i-1}=\cX{\ms_{i-1}}^*-\cX{\ms_{i-1}}\\
	-d_{i-1}(\cX{\ms_{i-1}}^*+x_{i-1})+d_{i-1}(\cX{\ms_{i-1}}^*)
	& \text{if }x_i=\cX{\ms_i}^*-\cX{\ms_i},\;x_{i-1}=\cX{\ms_{i-1}}-\cX{\ms_{i-1}}^*\\
	-d_{i-1}(\cX{\ms_{i-1}}^*-x_{i-1})+d_{i-1}(\cX{\ms_{i-1}}^*)
	& \text{if }x_i=\cX{\ms_i}^*-\cX{\ms_i},\;x_{i-1}=\cX{\ms_{i-1}}^*-\cX{\ms_{i-1}}.
	\end{cases}
	\end{align*}
	Since all the $c_i$ are monotone strictly increasing, we have that 
	\begin{align*}
	&a_i(x_i)\quad\begin{cases}
	>0 & x_i>0\\
	= 0 & x_i = 0\\
	<0 & x_i<0.
	\end{cases}, &
	&\forall i\in\{2,3,\dots,N\}.
	\end{align*}
	We have already selected $x_1$ and $x_N$ according to \eqref{eq:1-N}
	to obtain \eqref{eq:ab-1-N}.  Our goal is now to select the remaining $x_i$,
	$i\in\{2,3,\dots,N-1\}$ according to \eqref{eq:plus} or
	\eqref{eq:minus} so that we also have
	\begin{align*}
	&b_{i-1}(x_{i-1})\quad\begin{cases}
	>0 & x_{i-1}>0\\
	= 0 & x_{i-1} = 0\\
	<0 & x_{i-1}<0,
	\end{cases}
	&\forall i\in\{2,3,\dots,N\}.
	\end{align*}
	which would require us to have $\forall i\in\{2,3,\dots,N\}$
	\begin{align*}
	&\begin{cases}
	\text{$d_{i-1}$ monotone strictly increasing}
	& \text{if }x_i=\cX{\ms_i}-\cX{\ms_i}^*,\;x_{i-1}=\cX{\ms_{i-1}}-\cX{\ms_{i-1}}^*\\
	& \text{or } x_i=\cX{\ms_i}^*-\cX{\ms_i},\;x_{i-1}=\cX{\ms_{i-1}}^*-\cX{\ms_{i-1}}\\
	\text{$d_{i-1}$ monotone strictly decreasing}
	& \text{if }x_i=\cX{\ms_i}-\cX{\ms_i}^*,\;x_{i-1}=\cX{\ms_{i-1}}^*-\cX{\ms_{i-1}}\\
	& \text{or } x_i=\cX{\ms_i}^*-\cX{\ms_i},\;x_{i-1}=\cX{\ms_{i-1}}-\cX{\ms_{i-1}}^*
	\end{cases}
	\end{align*}
	It turns out that this is always possible because there is an even
	number of the $d_{i-1}$ with $i\in\{2,3,\dots,N\}$ are monotone
	strictly decreasing (recall that $d_N$ is monotone strictly
	decreasing and there are in total an odd number of $d_i$ that are monotone
	strictly decreasing). All we need to do is to start with $x_1$ as in
	\eqref{eq:1-N} and alternate between \eqref{eq:plus} and
	\eqref{eq:minus} each time $d_{i-1}$ is monotone strictly
	decreasing. Since there is an even number of the $d_{i-1}$ with
	$i\in\{2,3,\dots,N\}$, we will end up with $x_N$ as in
	\eqref{eq:1-N}.
	
	\medskip
	
	The coordinate transformation constructed above, leads us to a system
	of the following form
	\begin{subequations}
		\label{eq:arcakform}
		\begin{align}
		\dot x_1&=-a_1(x_1)-b_N(x_N)\\
		\dot x_2&=-a_2(x_2)+b_1(x_1)\\
		&\;\vdots\\
		\dot x_N&=-a_N(x_N)+b_{N-1}(x_{N-1})
		\end{align}
	\end{subequations}
	To prove the local stability result, we apply the secant criterion
	\cite{tyson1978dynamics,thron1991secant,arcak2006diagonal} to the
	local linearization of this system around the equilibrium $x_i=0$,
	$\forall i$, which has a Jacobian matrix of the form
	\begin{align}\label{eq:Jacobian}
	\matt{ -\PDeriv{a_1(x_1)}{x_1}\Big|_{x_1 = 0}&0&\cdots&0&-\PDeriv{b_N(x_N)}{x_N}\Big|_{x_N = 0}\\
		\PDeriv{b_1(x_1)}{x_1}\Big|_{x_1 = 0}& -\PDeriv{a_2(x_2)}{x_2}\Big|_{x_2 = 0}&\ddots&&0\\
		0&\PDeriv{b_2(x_2)}{x_2}\Big|_{x_2 = 0}&-\PDeriv{a_3(x_3)}{x_3}\Big|_{x_3 = 0}&\ddots&\vdots\\
		\vdots&\ddots&\ddots&\ddots&0\\
		0&\cdots&0&\PDeriv{b_{N-1}(x_{N-1})}{x_{N-1}}\Big|_{x_{N-1} = 0}&-\PDeriv{a_N(x_N)}{x_N}\Big|_{x_N = 0}},
	\end{align}
	where
	\begin{align*}
	\PDeriv{a_i(x_i)}{x_i}\Big|_{x_i = 0} &=\PDeriv{c_i(s_i)}{s_i}\big|_{s_i=\cX{\ms_i}^*} >0 \\
	\PDeriv{b_{i}(x_{i})}{x_{i}}\Big|_{x_{i} = 0} &= \begin{cases}
	\PDeriv{d_i(s_i)}{s_i}\big|_{s_i=\cX{\ms_i}^*}>0 & \text{if } d_{i}\text{ monotone increasing}\\
	-\PDeriv{d_i(s_i)}{s_i}\big|_{s_i=\cX{\ms_i}^*}>0 & \text{if } d_{i}\text{ monotone decreasing,}
	\end{cases}
	&\forall i\in\{1,2,\dots,N\}.
	\end{align*}
	This matrix matches precisely the one considered in the secant
	criteria, which states that the Jacobian matrix \eqref{eq:Jacobian}
	is Hurwitz if \eqref{eq:cond1} holds.
	
	\medskip
	
	For the global asymptotic stability result we use \cite[Corollary
	3]{arcak2006diagonal}, which applies precisely to systems of the
	form \eqref{eq:arcakform} with
	\begin{align*}
	x_i a_i(x_i)&>0, &
	x_i b_i(x_i)&>0, & \forall x_i\neq 0, i\in\{1,2,\dots,N\}.
	\end{align*}
	Two additional conditions are needed by \cite[Corollary
	3]{arcak2006diagonal}: 
	\begin{align}\label{eq:b-unbounded}
	\lim\limits_{|{x_i}|\to \infty}\int_{0}^{x_i}b_i(\sigma)d\sigma = \infty.
	\end{align}
	and there must exist $\phi_i>0$, $\forall i\in\{1,2,\dots,N\}$ for
	which
	\begin{subequations}
		\begin{align}
		\label{eq:a-secant}\frac{b_i(x_i)}{a_i(x_i)}&\le \phi_i, \quad\forall i,\;x_i\neq 0,\\
		\label{eq:b-secant}\prod_{i=1}^{N}\phi_i &< \sec\left({\frac{\pi}{N}}\right)^N.
		\end{align}
	\end{subequations}
	The first condition \eqref{eq:b-unbounded} holds because our
	functions $b_i$ are all zero at zero and monotone strictly
	increasing.
	
	We then prove two conditions to be sufficient for \eqref{eq:a-secant} to be satisfied. Before proceeding, we make the following observations:
	\begin{align}
	\label{eq:ratios}
	\frac{b_i(x_i)}{a_i(x_i)} &\eqdef \begin{cases}
	\frac{ d_{i}(\cX{\ms_{i}}^*+x_{i})-d_{i}(\cX{\ms_{i}}^*)}{c_i(\cX{\ms_i}^*+x_i)-c_i(\cX{\ms_i}^*)} \quad &\text{or }\quad \frac{-d_{i}(\cX{\ms_{i}}^*+x_{i})+d_{i}(\cX{\ms_{i}}^*)}{c_i(\cX{\ms_i}^*+x_i)-c_i(\cX{\ms_i}^*)}  \quad\text{if }  x_i=\cX{\ms_i}-\cX{\ms_i}^*\\\\
	\frac{d_{i}(\cX{\ms_{i}}^*-x_{i})-d_{i}(\cX{\ms_{i}}^*)}{ -c_i(\cX{\ms_i}^*-x_i)+c_i(\cX{\ms_i}^*)} \quad &\text{or }\quad \frac{-d_{i}(\cX{\ms_{i}}^*-x_{i})+d_{i}(\cX{\ms_{i}}^*)}{ -c_i(\cX{\ms_i}^*-x_i)+c_i(\cX{\ms_i}^*)} \quad\text{if } x_i=\cX{\ms_i}^*-\cX{\ms_i}.
	\end{cases}
	\end{align}
	\begin{align}
	\label{eq:derivai}
	\PDeriv{a_i(x_i)}{x_i} &\eqdef \begin{cases}
	\frac{\partial}{\partial{x_i}}c_i(\cX{\ms_i}^*+x_i) \qquad& \text{if }  x_i=\cX{\ms_i}-\cX{\ms_i}^*\\
	-\frac{\partial}{\partial{x_i}}c_i(\cX{\ms_i}^*-x_i) \qquad&  \text{if }  x_i=\cX{\ms_i}^*-\cX{\ms_i}.
	\end{cases}
	\end{align}
	\begin{align}
	\label{eq:derivbi}
	\PDeriv{b_i(x_i)}{x_i} &\eqdef \begin{cases}
	\frac{\partial}{\partial{x_i}}d_{i}(\cX{\ms_{i}}^*+x_{i}) \quad &\text{or } \quad -\frac{\partial}{\partial{x_i}}d_{i}(\cX{\ms_{i}}^*+x_{i}) \qquad \text{if }  x_i=\cX{\ms_i}-\cX{\ms_i}^*\\
	\frac{\partial}{\partial{x_i}}d_{i}(\cX{\ms_{i}}^*-x_{i}) \quad &\text{or }\quad -\frac{\partial}{\partial{x_i}}d_{i}(\cX{\ms_{i}}^*-x_{i})\qquad \text{if }  x_i=\cX{\ms_i}^*-\cX{\ms_i}.
	\end{cases}
	\end{align}
	
	First, we prove that the condition 
	\begin{align}
	\label{eq:gas}
	\Big|\frac{d_i(z_i) - d_i(\cX{\ms_i}^*)}{c_i(z_i) - c_i(\cX{\ms_i}^*)} \Big| \leq \phi_i \quad \forall z_i \neq \cX{\ms_i}^*.
	\end{align}
	implies \eqref{eq:a-secant}. We see that \eqref{eq:gas} implies that
	
	\begin{align}
	\label{eq:expand}\frac{d_i(z_i) - d_i(\cX{\ms_i}^*)}{c_i(z_i) - c_i(\cX{\ms_i}^*)}\leq \phi_i\quad\text{and }\quad -\frac{d_i(z_i) - d_i(\cX{\ms_i}^*)}{c_i(z_i) - c_i(\cX{\ms_i}^*)}\leq \phi_i\qquad \forall z_i \neq \cX{\ms_i}^*.
	\end{align}
	With the change of co-ordinates $x_i = -\cX{\ms_i}^* + z_i$ and $x_i = \cX{\ms_i}^* - z_i$, we see that \eqref{eq:expand} implies that
	\begin{align}
	\label{eq:coordchange}
	\begin{split}
	\frac{ d_{i}(\cX{\ms_{i}}^*+x_{i})-d_{i}(\cX{\ms_{i}}^*)}{c_i(\cX{\ms_i}^*+x_i)-c_i(\cX{\ms_i}^*)}\leq \phi_i \: &\text{, }\quad \frac{-d_{i}(\cX{\ms_{i}}^*+x_{i})+d_{i}(\cX{\ms_{i}}^*)}{c_i(\cX{\ms_i}^*+x_i)-c_i(\cX{\ms_i}^*)}\leq \phi_i \\
	\frac{d_{i}(\cX{\ms_{i}}^*-x_{i})-d_{i}(\cX{\ms_{i}}^*)}{ -c_i(\cX{\ms_i}^*-x_i)+c_i(\cX{\ms_i}^*)}\leq \phi_i \: &\text{, }\quad \frac{-d_{i}(\cX{\ms_{i}}^*-x_{i})+d_{i}(\cX{\ms_{i}}^*)}{ -c_i(\cX{\ms_i}^*-x_i)+c_i(\cX{\ms_i}^*)}\leq \phi_i \qquad  \forall x_i \neq 0.
	\end{split}
	\end{align}
	From \eqref{eq:ratios}, we conclude that \eqref{eq:coordchange} implies \eqref{eq:a-secant}.
	
	We then prove that the condition
	\begin{align}
	\label{eq:gas2}
	\Big|\PDeriv{d_i(z_i)}{z_i}\Big| \leq \phi_i\PDeriv{c_i(z_i)}{z_i},\quad z_i \neq \cX{\ms_i}^*
	\end{align}
	implies \eqref{eq:a-secant}. We see that \eqref{eq:gas2} implies that
	\begin{align}
	\label{eq:expand2}\PDeriv{d_i(z_i)}{z_i} \leq \phi_i\PDeriv{c_i(z_i)}{z_i}\quad\text{and }\quad -\PDeriv{d_i(z_i)}{z_i} \leq \phi_i\PDeriv{c_i(z_i)}{z_i}\qquad \forall z_i \neq \cX{\ms_i}^*
	\end{align}
	With the change of co-ordinates $x_i = -\cX{\ms_i}^* + z_i$ and $x_i = \cX{\ms_i}^* - z_i$, \eqref{eq:expand2} can be seen to imply that
	\begin{align}
	\label{eq:coordchange2}
	\begin{split}
	\frac{\partial}{\partial{x_i}}d_{i}(\cX{\ms_{i}}^*+x_{i})\leq \phi_i\frac{\partial}{\partial{x_i}}c_i(\cX{\ms_i}^*+x_i) \: &\text{, }\: -\frac{\partial}{\partial{x_i}}d_{i}(\cX{\ms_{i}}^*+x_{i})\leq \phi_i\frac{\partial}{\partial{x_i}}c_i(\cX{\ms_i}^*+x_i) \\
	\frac{\partial}{\partial{x_i}}d_{i}(\cX{\ms_{i}}^*-x_{i})\leq -\phi_i\frac{\partial}{\partial{x_i}}c_i(\cX{\ms_i}^*-x_i) \: &\text{, }\: -\frac{\partial}{\partial{x_i}}d_{i}(\cX{\ms_{i}}^*-x_{i})\leq -\phi_i\frac{\partial}{\partial{x_i}}c_i(\cX{\ms_i}^*-x_i) \quad  \forall x_i \neq 0
	\end{split}
	\end{align}
	From \eqref{eq:derivai}--\eqref{eq:derivbi}, we can observe that \eqref{eq:coordchange2} implies that
	\begin{align}
	\label{eq:ab}
	\PDeriv{b_i(x_i)}{x_i}<\phi_i\PDeriv{a_i(x_i)}{x_i} \quad \forall x_i \neq 0 
	\end{align}
	Let $h_i(x_i) \eqdef b_i(x_i) - \phi_i a_i(x_i)$. Then, \eqref{eq:ab} implies that
	\begin{align}
	\label{eq:h}
	\PDeriv{h_i(x_i)}{x_i}\leq 0  \quad \forall x_i \neq 0.
	\end{align}
	Since $a_i(0) = 0$ and $b_i(0) = 0$, we know that $h_i(0) = 0$. Therefore \eqref{eq:h} implies that 
	\begin{align*}
	h_i(x_i)\begin{cases}
	\leq 0 &\forall x_i >0\\
	\geq 0 &\forall x_i <0,
	\end{cases}
	\end{align*}
	which further implies that
	\begin{align}
	\label{eq:penultimate}
	b_i(x_i) \begin{cases}
	\leq \phi_ia_i(x_i) &\forall x_i >0\\
	\geq \phi_ia_i(x_i) &\forall x_i <0\\
	\end{cases}
	\end{align}
	Since   
	\begin{align*}
	a_i(x_i)\quad\begin{cases}
	>0 & x_i>0\\
	< 0 & x_i < 0\\
	\end{cases}
	\end{align*}
	\eqref{eq:penultimate} implies \eqref{eq:a-secant}, hence completing our proof. \frQED
\end{proof-theorem}

\draftbreak

\begin{corollary}
	\label{cor:suff}
	Consider a Repressilator network, that consists of an odd number $N$
	of equal single-gene TR repressor modules ($F=1$) connected in
	feedback as in Figure~\ref{fig:fbk_spc}, with $q=2$. Under
	Assumption~\ref{as:tr-timescale} for the TR repressor modules, the
	network has a unique equilibrium point that is LAS if
	\begin{equation}
	\label{eq:lasrep}
	\ \frac{2\ptot \alpha \beta \frac{u^*}{\bbeta} }{\bar{\beta}K\gamma \left(1 + \frac{1}{K}(\frac{u^*}{\bbeta})^2\right)^2} < \sec\left(\frac{\pi}{2N}\right)^2,
	\end{equation}
	where $u^*$ is the unique solution to
	\begin{align}
	\label{eq:f1}
	\frac{\alpha\beta P^\mrm{tot}}{\gamma\Big(1 + \frac{(\frac{u^*}{\bbeta})^2}{K}\Big)} = u^*,
	\end{align}
	and is GAS if
	\begin{equation}
	\label{eq:gasrep}
	\frac{3\ptot \alpha\beta}{8\bar{\beta}\gamma}\sqrt{\frac{3}{K}}< \sec\left(\frac{\pi}{2N}\right)^2.
	\end{equation}
\end{corollary}

\begin{proof-corollary}{\ref{cor:suff}}
	
	We can apply Theorem $\ref{thm:secdiag}$ to analyze a Repressilator network, which consists of $N$ SISO repressor modules connected in negative feedback, with $q=2$. Each SISO repressor module can be further decomposed into two modules $\mv_1$ and $\mv_2$ where

	\begin{equation*}
	\begin{aligned}
	&\bm{{\mv}_1}\\
	\dotcX{\ms_0} &= u_1 - \bar{\beta} \cX{\ms_0}\\
	y_1 &\eqdef h(\cX{\ms_0}) = \begin{cases}
	\frac{\alpha P^{\tu{tot}}}{1 + \frac{1}{K}\cX{\ms_0}^2} \quad& \text{ if } \cX{\ms_0} \geq 0\\
	\alpha P^{\tu{tot}} \frac{1 + \frac{2}{K}\cX{\ms_0}^2}{1 + \frac{1}{K}\cX{\ms_0}^2} \quad& \text{ if } \cX{\ms_0} < 0.
	\end{cases}
	\end{aligned}\qquad\qquad
	\begin{aligned}
	&\bm{{\mv}_2}\\
	\dotcX{mRNA_1} &= u_2 - \gamma \cX{mRNA_1}\\
	y_2 &= \beta\cX{mRNA_1}
	\end{aligned}
	\end{equation*}
	It can be seen that a small modification has been made to $y_1$, the repressing output from the TR repressor module. Since our network is positive, this modification has no effect on the network behavior. However, this change makes it more straightforward to apply Theorem~\ref{thm:secdiag} to analyze this network, since the theorem relied on each output function being monotone strictly increasing or monotone strictly decreasing $\forall \cX{\ms_0}$.
	
	From the first part of Theorem \ref{thm:secdiag}, \eqref{eq:lasrep}--\eqref{eq:f1} guarantee that the equilibrium point of the Repressilator network will be LAS.
	
	For GAS, we first need to pick $\phi_1$ and $\phi_2$ to satisfy
	\begin{align*}
	\Big|\frac{\partial h(z)}{\partial z}\Big|&\leq \phi_1\bar{\beta} \quad z \neq \cX{\ms_0}^*\\
	\beta &\leq \phi_2 \gamma.
	\end{align*}
	It is straightforward to show that
	\begin{equation*}
	\max_{z\neq \cX{\ms_0}^*}\Big|\frac{\partial h(z)}{\partial z}\Big|=  \frac{3\alpha\ptot}{8}\sqrt{\frac{3}{K}},
	\end{equation*}
	so we can pick 
	\begin{equation*}
	\phi_1 = \frac{3\alpha\ptot}{8\bar{\beta}}\sqrt{\frac{3}{K}}\tu{,}\quad
	\phi_2 = \frac{\beta}{\gamma}.
	\end{equation*}
	From Equation \eqref{eq:a-secant} of Theorem \ref{thm:secdiag}, we can infer that \eqref{eq:gasrep} guarantees that the equilibrium point of the Repressilator network will be GAS.\frqed
\end{proof-corollary}

\bibliography{mybib}

\end{document}